\documentclass[journal,comsoc]{IEEEtran}

\usepackage[T1]{fontenc}
\usepackage{color}

\ifCLASSINFOpdf

\else

\fi

\usepackage{amsmath}

\usepackage[cmintegrals]{newtxmath}

\usepackage{amssymb,amsmath,graphicx}
\usepackage{float}
\usepackage{multirow}
\usepackage{multicol}
\newtheorem{definition}{Definition}
\newtheorem{theorem}{Theorem}

{\bfseries}{}
\usepackage{cite}
\usepackage{color}
\usepackage{comment}
\usepackage{mathtools}


\usepackage{etoolbox}

\AtBeginDocument{%
 \mathchardef\mathcomma\mathcode`\,
 \mathcode`\,="8000
}
{\catcode`,=\active
 \gdef,{\mathcomma\discretionary{}{}{}}
}

\makeatletter
\newcommand{\raisemath}[1]{\mathpalette{\raisem@th{#1}}}
\newcommand{\raisem@th}[3]{\raisebox{#1}{$#2#3$}}
\makeatother

\numberwithin{mysubcase}{mycase}

\definecolor{codegreen}{rgb}{0,0.6,0}
\definecolor{codegray}{rgb}{0.5,0.5,0.5}
\definecolor{codepurple}{rgb}{0.58,0,0.82}
\definecolor{backcolour}{rgb}{0.95,0.95,0.92}

\definecolor{bluekeywords}{rgb}{0.13,0.13,1}
\definecolor{greencomments}{rgb}{0,0.5,0}
\definecolor{redstrings}{rgb}{0.9,0,0}

\restylefloat{figure}

\newcommand{\eg}{e.g.\ }
\newcommand{\ie}{i.e.\ }
\newcommand{\cf}{cf.\ }
\newcommand{\etal}{ {\em et al.\,}}

\newcommand{\sw}[1]{\textcolor{blue}{#1}}
\newcommand{\swst}[1]{\textcolor{blue}{\st{#1}}}

\newcommand{\jh}[1]{\textcolor{red}{#1}}

\makeatletter
\def\@seccntformatinl#1{\csname the#1dis\endcsname\hskip 1em\relax}
\makeatother

\newlength\figwidth
\newcommand{\inserttabd}[4]{%
	\begin{table*}[!t]
		\begin{minipage}{\textwidth}
			\caption{#3}
			\label{#1}
			\centering
			\renewcommand{\arraystretch}{1.15}
			\begin{tabular}{#2}
				\hline
				#4
				\hline
			\end{tabular}
		\end{minipage}
\end{table*}}

\begin{document}

\title{Anonymous Single Sign-on with Proxy Re-Verification}

\author{Jinguang~Han,~\IEEEmembership{Senior Member,~IEEE,}
        Liqun~Chen,~\IEEEmembership{Member,~IEEE,}
        Steve~Schneider, Helen~Treharne, Stephan~ Wesemeyer ~and~ Nick~Wilson 
\thanks{\IEEEcompsocthanksitem J. Han, L. Chen, S. Schneider, H. Treharne  and S. Wesemeyer are  
     affiliated to the Surrey Centre for Cyber Security,  University of Surrey, Guildford, Surrey, GU2 7XH, United Kingdom\protect\\
E-mail: \{j.han, liqun.chen, s.schneider, h.treharne,  
s.wesemeyer\}@surrey.ac. uk
}
\thanks{N. Wilson is a Technical Architect, Rail Delivery Group, United Kingdom\protect\\
E-mail: nick.wilson@raildeliverygroup.org
}
}

\markboth{}%
{}

\maketitle

\begin{abstract}%
An anonymous Single Sign-On (ASSO)  scheme allows users to access multiple services anonymously using one credential.   We
propose a new ASSO scheme, where users can access services anonymously through the use of
anonymous credentials and unlinkably through the provision of designated
verifiers. Notably, verifiers cannot link a user's service requests even if they
collude. The novelty is that when a designated verifier is unavailable, a
central authority can authorise new verifiers to authenticate the user on behalf
of the original verifier. Furthermore, if required, a central verifier is
authorised to de-anonymise users and trace their service requests. We formalise
the scheme along with a security proof and provide an empirical evaluation of
its performance.  
This scheme can be applied to smart ticketing where
 minimising the collection of personal information of users is increasingly
important to transport organisations due to privacy regulations such as  General Data Protection Regulations (GDPR). 
\end{abstract}


\IEEEpeerreviewmaketitle

\section{Introduction}\label{sec:introduction}

\IEEEPARstart{S}{ingle} Sign-on (SSO) is a mechanism that enables a user to
access multiple services using only one credential. Existing SSO
solutions include OpenID \cite{recordon2006openid}, SAML \cite{cmj:saml2015},
and Kerberos \cite{2017MITKerberos}, {\em etc}. SSO systems can reduce a
user's burden on maintaining authentication credentials.

In order to protect users' privacy, anonymous SSO (ASSO) systems were
proposed in
\cite{2008ElmuftiAnonAuth,2010hanDynSSO,2013WangAnonSSO,2015LeeAnonSSO}. In
these systems,  a user's personal identifiable information (PII) was
considered, but the unlinkability of the user's service requests was not. 
Recently, Han {\em et al.}~\cite{hcsts:asso2018} proposed a
new ASSO scheme which protects the identity of both the user and her service
requests. Their scheme allows users to obtain a ticket from a ticket
issuer to access multiple intended services. The ticket consists of a set of
authentication tags that can only be validated by designated verifiers. %
Designated verifiers can validate their corresponding tags and
cannot link a user's service requests, even if they collude. A third
party, referred to as a central verifier, can de-anonymise a user's identity 
and trace her
service requests. 

In a transport application a ticket could represent an intended route of travel
(\eg from A to B to C). Traditionally, in the rail industry, tickets were paper
based and hence anonymous. In the context of smart ticketing, which is one of
the main digital strategies of the UK rail industry \cite{RTS2012}, customers'
data may be stored when buying tickets. Thus, it will be important to consider
passenger privacy in order to minimise the collection of personal information to
reflect the requirements of the recently introduced General Data Protection
Regulations (GDPR) \cite{EGDPR2016}. Nonetheless, a smart ticketing solution 
will still need to provide guarantees as to who owns and uses a rail ticket. 
Using an anonymous scheme such as Han\etal~\cite{hcsts:asso2018} means that 
passenger information leakage between different companies is prevented because 
each train operating company is considered to be a separate designated verifier.
However, the inclusion of a central verifier allows the relevant
transport authorities to identify passengers and their journeys. This is
important in the case of an emergency to enable transport authorities to know
who the passengers using their transport systems are. It could also provide
guards on a train access a user's whole journey information in order to provide
the best journey advice during travel if appropriate.

In \cite{hcsts:asso2018}, an authentication tag can only be validated by a
designated service provider, hence a user cannot access the services if the
service provider is off-line or unavailable. %
In a cloud environment and when a service provider is off-line, a user would
expect to be redirected to an alternative provider offering a similar service. 
While for a transport application (in the case of disruption), a ticket should
still be valid and authorised for use on a redirected route. For example, a
journey from A to C via B could be redirected to go via D and/or E when B is
disrupted. In such cases a user should not be required to buy or change her
ticket in order to access the alternative route. Moreover, in practice, the entities who hold the disruption information are disconnected from those who sell tickets. Therefore,
rail authorities and train companies should manage and be responsible for the redirected travel routes and disruption information with minimal impact on users.

In this paper, we propose a new ASSO scheme which extends the scheme presented
in \cite{hcsts:asso2018} to allow a central authority to authorise another
verifier to act as a proxy  and validate the authentication tags for a service
provider that is unavailable.  In the ticket scenario it thus provides a central
authority with the ability to allow a proxy verifier to validate a user's
ticket. Hence, proxy re-verification does not increase a user's authentication
burden in case of a disruption, \ie a user does not need to change her ticket.
Our new scheme also preserves the following features from the original scheme of
Han\etal~\cite{hcsts:asso2018}:
\begin{enumerate}%
\item{\em Multiple Access:} a user can use one ticket to access
multiple distinct services;%
\item{\em Anonymity:} a user can obtain a ticket from a ticket issuer without
releasing anything about her  PII to the ticket
issuer, especially, the ticket issuer cannot determine whether two ticket
are issued to the same user or two different users; %
\item{\em Unlinkability:} a designated verifier can determine whether a user is
authorised to access its service but cannot link a user's different service
requests nor collude with other verifiers to link a user's service requests;%
\item{\em Unforgeability:} tickets can only be issued by ticket issuers and
cannot be forged by other parties even the central authority; %
\item{\em Traceability:}  only the central verifier can de-anonymise a user and
trace the identities of the verifiers whose services the user is authorised to
access; %
\item{\em Double Spending Detection:} designated verifiers can detect and 
prevent a user
from making two authentication requests using the same authentication tag
but cannot de-anonymise the user;%
\end{enumerate}

\noindent{\em Contributions:} Our main contributions in this paper are 
summarised as follows: (1) an ASSO with proxy re-verification
 scheme providing the above features is formally constructed;  (2) 
the definition and security model are formalised;  (3) the scheme has been 
implemented and an empirical efficiency analysis is presented; 
(4) the security of our scheme is formally reduced to well-known complexity 
assumptions. 

The novelty of this paper is to prevent information leakage across multiple
verifiers and implement proxy re-verification. To the best of our knowledge, our
scheme is the first scheme to support users anonymously and unlinkably
authenticating to multiple service providers and allowing authorised proxy
verifiers to verify authentication on behalf of an original designated verifier
when that verifier is unavailable.

\subsection{Related Work}

In this subsection, we review the work which is most closely related to our 
scheme.  Previous authentication schemes mainly address the anonymity of users 
and implement multiple authentications using one credential.

\subsubsection{Anonymous Single-Sign-On schemes}

Elmufti {\em et al.} \cite{2008ElmuftiAnonAuth}  proposed an ASSO scheme which
is suitable to the Global System for Mobile communication (GSM). In
\cite{2008ElmuftiAnonAuth}, to access a service, a user needs to generate a new
one-time identity and uses it to authenticate to a trusted third party (TTP). If
the authentication is successful, the TTP forwards the user's one-time identity
to the service provider who provides the service. As a result, the service
provider cannot infer the user's real identity from this one-time
identity. However, in our scheme, users can authenticate to service providers
directly without the need of a TTP.

Han {\em et al.} \cite{2010hanDynSSO} proposed a generic construction of dynamic
SSO schemes where digital signature, broadcast encryption and zero-knowledge
proof are adopted. In  \cite{2010hanDynSSO}, after registering with the
system, a user obtains a credential which is the encryption of a signature
generated by the central authority on a set of service selected by the user and
her public key. Consequently, only the service providers whose services
have been selected by the user can decrypt the ciphertext and validate the
signature. To prevent sharing a credential, a user needs to prove the knowledge
of her secret key corresponding to the public key included in the credential.
Hence,  a user is anonymous only to the service providers who are not included
in the credential. Nevertheless, unlike in our scheme, service providers
 know the user's identity (public key) and link her 
service requests.

Wang {\em et al.} \cite{2013WangAnonSSO} proposed an ASSO scheme based on group
signatures \cite{bmw:gs2003}. When registering to the central authority, a
user  is issued a group member key. Then, to access a
service, a user generates a group signature by using her group member key. A
service provider checks whether the user is authorised to access services by
validating the correctness of the signature. Furthermore, the central authority
can use the open algorithm in the group signature scheme to trace a user's
identity. Notably, a user can access all services in the system, while in our
scheme a user can only access the selected services.

Lee \cite{2015LeeAnonSSO} proposed an efficient ASSO scheme based on Chebyshev
Chaotic Maps. When joining the system, an issuer (the smart card processing
center) issues temporary secret keys to users and service providers. To access a
service, a user interacts with a service provider to generate a session key by
using their respective temporary secret keys. A service request is granted if
and only if the  session key can be generated correctly; otherwise, the request
is denied. However, unlike our scheme, each service provider knows the identity
of the user accessing his service. Hence, multiple service providers can
profile a user's service requests if they collude. Moreover, a user can
again access all services in the system, while in our scheme a user can
only access the selected services.

\subsubsection{Proxy Re-Encryption} Mambo and Okamoto \cite{mo:pre1997}
introduced the definition of proxy cryptosystems that enable a delegator to
delegate the decryption power to a delegatee. Later, Blaze \cite{bbs:pre}
proposed an atomic proxy cryptography scheme where a  semi-trusted third party
called proxy can convert ciphertexts for one user  into ciphertexts for another
user if the third party is given a proxy key.

Shamir \cite{s:ibc1984} introduced an identity-based cryptosystem is a public
key cryptosystem where a user's public key can be any arbitrary string and her
secret key is obtained from a trusted central authority. Boneh and Franklin
\cite{bf:ibe2001} first proposed a practical identity-based encryption (IBE)
scheme based on paring.  Green and Ateniese \cite{ga:ibpre2007}  introduced the
concept of identity-based proxy re-encryption (IBPRE)  where a proxy can convert
a ciphertext for the original decryptor to a ciphertext for a designated
decryptor if the proxy obtains a re-encryption key from the original decrytor.
Han {\em et al.} \cite{hsm:ibpre2014} classified IBPRE schemes into two types
according to the generation of re-encryption keys: (1) re-encryption keys are
generated by the trusted central authority \cite{m:ibpre2007,wwmo:ibpre2010};
(2) re-encryption keys are generated by the original decryptors
\cite{ga:ibpre2007,ct:ibpre2007}. In
\cite{m:ibpre2007,wwmo:ibpre2010,ga:ibpre2007,ct:ibpre2007}, given a
re-encryption key, a proxy can convert all ciphertexts for the original
decryptor to ciphertexts for the designated decryptor. The differences between
our scheme and IBPRE schemes are: (1) a proxy is not required; (2) a re-key only
enables a proxy verifier to validate tickets on behalf the original verifier in
a specified period, instead of all tags.

\subsubsection{Designated Verifier Schemes}

Jakobsson\etal\cite{jsi:1996} introduced a designated verifier signature (DVS)
scheme which is a digital signature scheme where a signature can only be
verified by a single designated verifier. Furthermore, the verifier cannot
convince others that a signature is from the real signer since the verifier
could have generated the signature by himself. Fan\etal\cite{FAN2012944} presented
an attribute-based DVS scheme where a signature can be verified by a group of
verifiers whose attributes satisfies specified values. In our scheme, we adopt
the high level concept of a designated verifier, \ie given a valid
authentication tag, only the corresponding designated verifier and the
authorised proxy verifiers can validate it. The main difference between these
DVS schemes \cite{jsi:1996,FAN2012944} and our scheme is that only the
designated verifiers can verify a signature in DVS schemes, while in our scheme,
everyone can verify a tag's signature generated by the ticket issuer but only
the designated verifier of the tag can determine for whom it was generated.

Kuchta\etal\cite{kssssm:dvs2018} proposed an identity-based strong designated
verifier group signature (ID-SDVGS) scheme that can provide the features of both
designated verifier signatures and identity-based group signatures. In this
scheme, all entities must obtain secret keys from a trusted third party referred
to as ``private key generator'' (PKG). When joining the group, each user obtains
a member credential from the group manager (GM). Then, a user can use her
credential to anonymously generate a signature which can only be verified by the
designated verifier and can be opened/de-anonymized by the GM. The verifier
cannot convince others that the signature is from the real signer since the
verifier can generate the signature by himself. However, in our scheme, only the
secret keys of ticket verifiers are issued by the central authority. The secret
keys of other entities including the ticket issuer, users and the central
verifier are generated by themselves. Authentication tags can only be generated by
the ticket issuer and its correctness can be publicly verified. Nevertheless,
other entities cannot know for whom a tag is generated except the designated
verifier.

\subsubsection{$k$-time Anonymous Authentication Schemes}

Anonymous authentication schemes  enable a user to authenticate to a verifier
without releasing her  PII to the verifier. 
To limit the authentication time, Teranishi {\em et al.} \cite{tfs:k-taa04}
proposed a $k$- time anonymous authentication ($k$-TAA) scheme where users register with a central authority and
obtain an anonymous credential. A verifier generates $k$ authentication tags. For
each access, a user proves to the verifier that she has obtained a valid
credential from the central authority and selects a fresh authentication tag. As
a result, no party can identify  a user if she authenticates no more than $k$
times, while any party can identify a user if she authenticates more than $k$
times. In  \cite{tfs:k-taa04}, the central authority decides a user's access
permission and service verifiers do not have control on the access permissions.


Camenisch {\em et al.} \cite{chklm:k-taa06} proposed a periodic $k$-TAA scheme
where a user can anonymously authenticate herself to a service verifier no more
than $k$ times in a given time period. The authentication tags automatically
refresh every time period. When a user makes an anonymous authentication
request, she proves to a verifier that she has obtained a valid credential (CL
signature~\cite{camenisch2002signature}) from the central authority.
Lastly, Camenisch {\em et al.} proposed an identity mixer scheme
\cite{camenisch2010,ibmmixer} in which users need to obtain a credential for
their attributes. To access a service, a user proves to the service verifier
that she has the required attributes.

In all these schemes
\cite{tfs:k-taa04,nr:k-taa05,chklm:k-taa06,camenisch2010,ibmmixer},
authentication is not bound to a particular verifier, whereas in our scheme an
authentication tag can only be verified by a designated verifier. Furthermore,
$k$-TAA schemes allow verifiers to de-anonymise a user's identity when she has
authenticated more than $k$ times, while in our scheme a service verifier can
detect whether a user has used the tag (double spending) but cannot de-anonymise
a user's identity. Notably, our scheme allows  a central verifier to
de-anonymise a user and trace her service requests. 

In Table \ref{comparison}, we compare our scheme with related 
ASSO schemes in terms of anonymity, the inclusion of a designated verifier, 
traceability, re-verification, whether a trusted third party (TTP) is required 
to authenticate users on behalf of service provers as well as efficiency which 
mainly considers whether bilinear groups are required or not. 

\begin{table*}\caption{The Comparison Between Our Scheme and Related Schemes}\label{comparison}
\centering
\begin{tabular}{|c|c|c|c|c|c|c|}
\hline
Schemes & Anonymity & Designated Verifiers & Traceability & Re-Verification & Trusted Third Party (TTP)  & Efficiency (bilinear group)\\
\hline
Elmufti {\em et al.} \cite{2008ElmuftiAnonAuth}  & $\checkmark$ & $\times$ & $\checkmark$ & $\times$ & $\checkmark$ & $\times$\\
\hline
Han {\em et al.} \cite{2010hanDynSSO} & $\times$ & $\times$ & $\checkmark$ & 
$\times$ & $\times$ &   not applicable\\
\hline
Wang {\em et al.} \cite{2013WangAnonSSO} & $\checkmark$ & $\times$ &  
$\checkmark$ & $\times$ & $\times$ &   not applicable\\
 \hline
Lee \cite{2015LeeAnonSSO} & $\checkmark$ & $\times$ & $\times$ & $\times$ & $\times$ & $\times$\\
 \hline
Han {\em et al.} \cite{hcsts:asso2018} & $\checkmark$ & $\checkmark$  & $\checkmark$  & $\times$ & $\times$ & $\checkmark$ \\
 \hline
 Our Scheme & $\checkmark$  & $\checkmark$  & $\checkmark$  & $\checkmark$  & $\times$ & $\checkmark$ \\
 \hline
\end{tabular}
\end{table*}

\subsection{Paper Organisation}%

The remainder of this paper is organised in the following sections.
Section~\ref{overview} provides a high-level overview of our scheme and its
security requirements. Section~\ref{def_security} introduces the formal
definition and security model. Section \ref{preli} presents the preliminaries
for our scheme and a formal construction of our scheme is given in
Section~\ref{const}. Section~\ref{security_analysis} and  Section~\ref{perf}
present the security proof and the performance evaluation of our scheme,
respectively. Finally, Section~\ref{sec:conc} concludes the paper. %


\section{Scheme Overview and  Security Properties}\label{overview}

The notation used throughout this paper is summarised in Table
\ref{syntax}.%

\begin{table*}[!h]\caption{Notation Summary}\label{syntax}
\centering
\begin{tabular}{|c|l|c|l|}
\hline
Notation & ~~~~~~~~~Explanations & Notation &~~~~~~~~~~~~~~~~~ Explanations \\
\hline
$1^{\ell}$   ~             &A security parameter  &$\mathcal{V}_{i}$ & The $i$-th ticket 
verifier\\
$\mathcal{CA}$ ~    &Central authority &$J_{U}$             ~    &The service 
set of $\mathcal{U}$ consisting of the  \\
$\mathcal{I}$ ~       &Ticket issuer  &  & identities of ticket verifiers \& $ID_{CV}$ \\
$\mathcal{V}$   ~     &Ticket verifier &$PP$ & Public parameters\\
$\mathcal{U}$     ~   &User   & $Ps_{U}$               &A set of pseudonyms of  
$\mathcal{U}$ \\
$\mathcal{CV}$~        &Central verifier  &$Ps_{V}$     ~     & The  pseudonym 
generated for  $\mathcal{V}$ \\
$ID_{I}$          ~      &The identity of $\mathcal{I}$ &$Tag_{V}$            
~   &  An  authentication tag for  $\mathcal{V}$\\
$ID_{V}$              ~  &The identity of $\mathcal{V}$ ~  
&$Tag_{CV}$             &   An authentication tag for $\mathcal{CV}$\\
$ID_{U}$                ~&The identity of $\mathcal{U}$  ~&$T_{U}$             
~  & A ticket issued to $\mathcal{U}$\\
$ID_{CV}$                ~&The identity of $\mathcal{CV}$  ~& 
$|X|$                    ~& The cardinality of the set $X$  \\
$\epsilon(\ell)$       ~&A negligible function in $\ell$  ~& 
$x\stackrel{R}{\leftarrow} X$ ~&$x$ is randomly selected from the set $X$\\
$\sigma_{I}$         ~&The credential of $\mathcal{I}$   ~& $A(x)\rightarrow y$ 
~&$y$ is computed by running the \\
$\sigma_{V}$         ~&The credential of $\mathcal{V}$   ~& & algorithm $A(\cdot)$  with input $x$\\
$\sigma_{U}$        ~&The credential of $\mathcal{U}$    ~& 
$\mathcal{KG}(1^{\ell})$ ~& A secret-public key pair generation  algorithm\\
$\sigma_{CV}$     ~    &The credential of $\mathcal{CV}$ &  $\mathcal{BG}(1^{\ell})$ ~&A bilinear group generation algorithm\\
$MSK$ &Master Secret Key & PPT & Probable polynomial-time \\
$H_1, H_2$, $H_{3}$ &Cryptographic hash functions & $p$ & A prime number\\
\hline
\end{tabular}
\end{table*}

Our ASSO with flexible verification scheme consists of the following entities: 
\begin{itemize}%
\item a trusted central authority, $\mathcal{CA}$, which initialises the system,
issues credentials to  other entities in the scheme and authorises proxy
verification;%
\item a  user, $\mathcal{U}$, who wants to access some distinct services
anonymously and unlinkabily; %
\item a ticket issuer, $\mathcal{I}$, issues tickets to registered, yet
anonymous users for a set of selected services; %
\item a designated  verifier, $\mathcal{V}$, who can only validates the
authentication tags generated for him and cannot link a user's service requests;%
\item an authentication tag, $Tag_V$, which is bound to a user $\mathcal{U}$ and
a designated verifier $\mathcal{V}$ and is used to convince $\mathcal{V}$ that
$\mathcal{U}$ is authorised to access its service;%
\item a ticket, $T_U$, which consists of a set of authentication tags generated
for the designated verifiers  of the requested services;%
\item a central verifier, $\mathcal{CV}$, which is another trusted third party
which, given a ticket $T_{U}$, can de-anonymise the identities of the user and
trace her service requests.%
\end{itemize}

\begin{figure*}[]
\centering
\includegraphics[width=14cm,height=8cm]{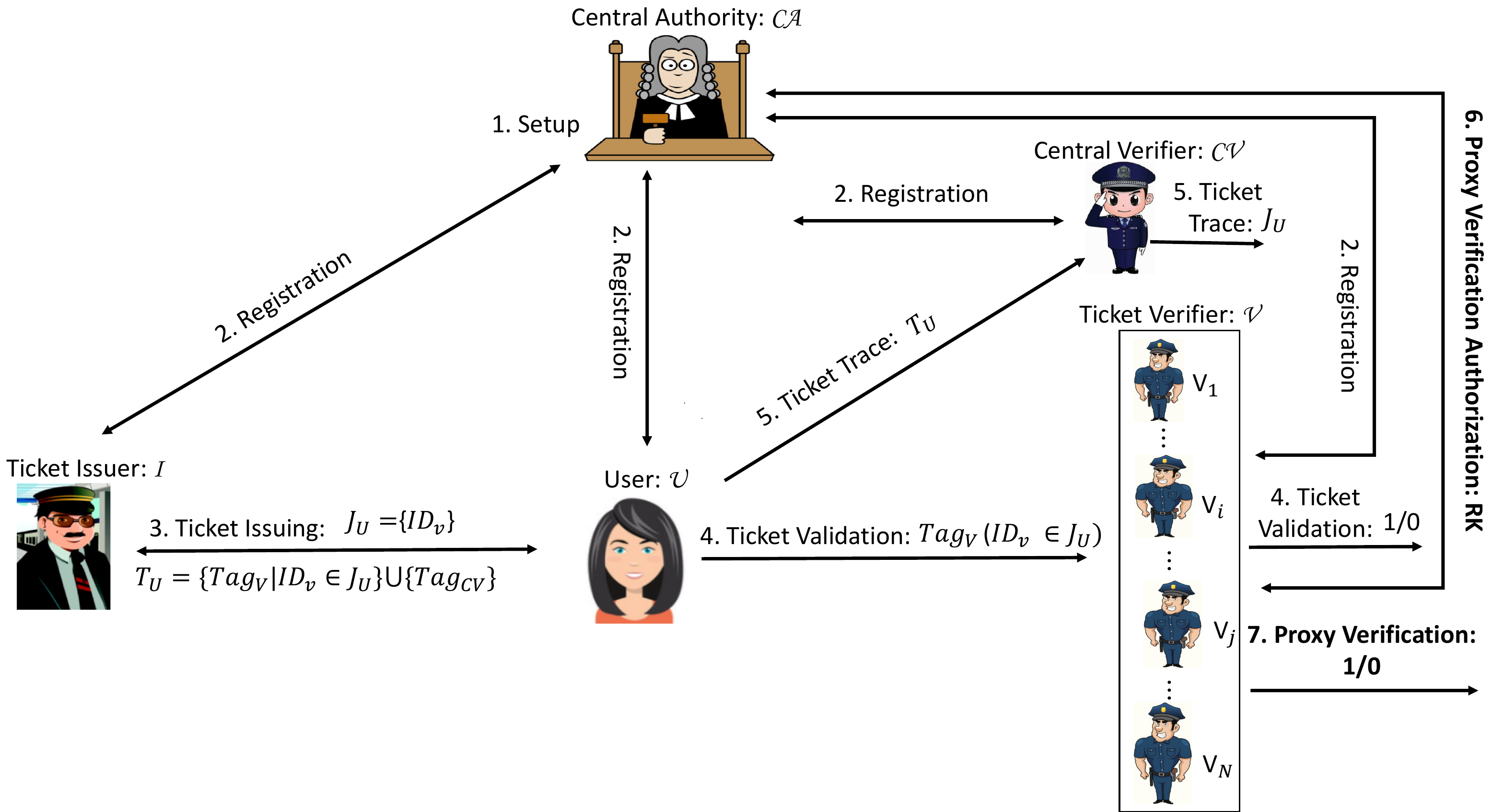}\caption{Pictorial
 description of our scheme}\label{fig:frame} 
\end{figure*}

\subsection{Overview of proposed scheme}%
A simplified pictorial description of our scheme is presented in Fig.
\ref{fig:frame}. $\mathcal{CA}$ initialises the system. When
joining the system,  $\mathcal{I}$,  $\mathcal{U}$,  $\mathcal{V}$ and   $\mathcal{CV}$  authenticate to the  $\mathcal{CA}$ and obtain their credentials from $\mathcal{CA}$. To buy a ticket, $\mathcal{U}$ sends her service information
$J_{U}$ consisting of a set of verifiers' identity $ID_{V}$ to $\mathcal{I}$.
Subsequently, $\mathcal{I}$ generates a ticket $T_{U}$ for $\mathcal{U}$. The
ticket comprises a set of tags $T_{U}=\{Tag_{V}|ID_{V}\in J_{U}\}\cup\{Tag_{CV}\}$ which can
only be validated by the corresponding designated verifiers. When being
validated by $\mathcal{V}$, $\mathcal{U}$ sends the corresponding tag $Tag_{V}$
to $\mathcal{V}$. In the case that $\mathcal{U}$'s service information needs to
be traced, $\mathcal{CV}$ is
allowed to  trace the whole service information of $\mathcal{U}$ given a ticket
$T_{U}$. Especially, when the original verifier $\mathcal{V}$ is unavailable, $\mathcal{CA}$ can authorise a new verifier $\mathcal{V}'$ to validate the tag on behalf of $\mathcal{V}$.

\subsection{Security Properties of Our Scheme}%

Having defined the different entities and described how they interact, we now 
list the security properties of our scheme:

\noindent{\em Anonymity:} a user can obtain a ticket from a ticket issuer
anonymously;

\noindent{\em Unlinkability:} a designated verifier cannot link a user's different
service requests nor collude with other verifiers to link a user's service
requests; %

\noindent{\em Unforgeability:} tickets are generated by ticket issuers and
cannot be forged by other parties even the central authority;%

\noindent{\em Traceability:} given a valid ticket, $\mathcal{CV}$ can
de-anonymise the ticket holder and trace her service requests;%

\noindent{\em Proxy Re-verification:} in the case that a designated verifier
$\mathcal{V}$ is unavailable, $\mathcal{CA}$ can assign one
or more verifiers $\mathcal{V}'$ to validate a user's tag designated for 
$\mathcal{V}$;

\noindent{\em Double Spending:} a designated verifier can detect whether a tag has
been used or not, but cannot de-anonymise the user.%


\section{Formal Definition and Security Requirement}\label{def_security}
In this section, we review the formal definition and security requirement of ASSO with proxy re-verification.

\subsection{Formal Definition}\label{formal_definition}

The definition of ASSO with proxy  re-verification is formalised by the following seven algorithms:

\medskip
\begin{enumerate}
\item{\sf Setup}
$(1^{\ell})\rightarrow \left(MSK,PP\right).$ Taking as input a security
parameter $1^{\ell}$, $\mathcal{CA}$ outputs a master secret key $MSK$ and the public
parameters $PP$.
\medskip

\item{\sf Registration}: This algorithm consists of the following sub-algorithms:\\
\noindent{\em Ticket-Issuer-Reg}\,%
$(\mathcal{I}(ID_{I},SK_{I},PK_{I},PP)\leftrightarrow\mathcal{CA}($ $MSK,PP))\rightarrow(\sigma_{I},(ID_{I},PK_{I})).$ This
algorithm is executed between $\mathcal{CA}$ and $\mathcal{I}$. $\mathcal{I}$
runs the secret-public key pair generation algorithm
$\mathcal{KG}(1^{\ell})\rightarrow(SK_{I},PK_{I})$, takes  its identity
$ID_{I}$, secret-public key pair $(SK_{I},PK_{I})$ and the public parameters
$PP$ as inputs, and outputs a credential $\sigma_{I}$. $\mathcal{CA}$  inputs
the master secret key $MSK$ and the public
parameters $PP$, and outputs $\mathcal{I}$'s identity and public key 
$(ID_{I},PK_{I})$.%
\medskip

\noindent{\em Ticket-Verifier-Reg}\,%
$(\mathcal{V}(ID_{V})\leftrightarrow\mathcal{CA}(MSK,PP))$
$\rightarrow((\sigma_{V},SK_{V}),(ID_{V},$ $ SK_{V}))$. This algorithm is
executed between $\mathcal{CA}$  and $\mathcal{V}$. $\mathcal{V}$ takes as 
input its identity $ID_{V}$, and outputs a credential $\sigma_{V}$ and a secret
key $SK_{V}$. $\mathcal{CA}$ takes as input the master secret key $MSK$, and
the public parameters $PP$, and outputs $\mathcal{V}$'s identity and secrete key
$(ID_{V},SK_{V})$.
\medskip

\noindent{\em User-Reg}\,%
$(\mathcal{U}(ID_{U},SK_{U},PK_{U},PP)\leftrightarrow
\mathcal{CA}(MSK,PP))\rightarrow(\sigma_{U},(ID_{U},PK_{U})).$ This
algorithm  is executed between $\mathcal{CA}$  and $\mathcal{U}$. $\mathcal{U}$
runs $\mathcal{KG}(1^{\ell})\rightarrow(SK_{U},PK_{U})$, takes as input its
identity $ID_{U}$, secret-public key pair $(SK_{U},PK_{U})$ and the public
parameters $PP$, and outputs a credential $\sigma_{U}$. $\mathcal{CA}$ takes as
input the master secret key $MSK$ and the
public parameters $PP$, and outputs $\mathcal{U}$'s identity and public key
$(ID_{U},PK_{U})$. 
\medskip

\noindent{\em Central-Verifier-Reg}\,%
$(\mathcal{CV}(ID_{CV},SK_{CV},PK_{CV},PP)$ $\leftrightarrow
\mathcal{CA}(MSK,PP))\rightarrow(\sigma_{CV},$ $(ID_{CV},PK_{CV})).$
This  algorithm is executed between $\mathcal{CA}$  and $\mathcal{CV}$.
$\mathcal{CV}$ runs $\mathcal{KG}(1^{\ell})\rightarrow(SK_{CV},PK_{CV})$, takes
as input its identity $ID_{CV}$, secret-public key pair $(SK_{CV},PK_{CV})$ and
the public parameters $PP$, and outputs a credential $\sigma_{CV}$.
$\mathcal{CA}$ takes as input the master secret key $MSK$ and the public parameters $PP$, and outputs
$\mathcal{CV}$'s identity and public key $(ID_{CV},PK_{CV})$. 
\medskip

\item{\sf Ticket-Issuing}\,%
$(\mathcal{U}(J_{U},SK_{U},PK_{U},\sigma_{U},PP)\leftrightarrow\mathcal{I}$
$(SK_{I},PK_{I},PP)\rightarrow (T_{U},J_{U}).$ This  algorithm is executed
between  a user $\mathcal{U}$ and a ticket issuer $\mathcal{I}$. $\mathcal{U}$
inputs her service information $J_{U}$ which consists of the identities of
ticket verifiers for the services which she wants to access as well as the identity of
the central verifier $ID_{CV}$,  her secret-public key pair $(SK_{U},PK_{U})$,
her credential $\sigma_{U}$  and the public parameters $PP$ and outputs a ticket
$T_{U}=\{Tag_{V}|ID_{V}\in J_{U}\}\cup \{Tag_{CV}\}$ where the authentication
tags $Tag_{V}$ and $Tag_{CV}$ can only be validated by the designated verifier
$\mathcal{V}$ with $ID_{V}\in J_{U}$ and the central verifier $\mathcal{CV}$,
respectively. $\mathcal{I}$ takes as input his secret-public key pair
$(SK_{I},PK_{I})$, the public parameters $PP$ and outputs the service
information $J_{U}$. 
\medskip
\item{\sf Ticket-Validation}%
$(\mathcal{U}(SK_{U}, PK_{U}, Tag_{V}, PP) \leftrightarrow \mathcal{V}(($ $ID_{V},
SK_{V}), PK_{I}, PP)) \rightarrow (\bot, (1, Tag_{V})/(0, Tag_{V}))$. This is an
interactive algorithm and is executed between $\mathcal{U}$ and $\mathcal{V}$
with $ID_{V}\in J_{U}$. $\mathcal{U}$ takes as input her secret-public key pair
$(SK_{U},PK_{U})$,  the authentication tag $Tag_{V}$ and the public parameters
$PP$, and outputs $\bot$. $\mathcal{V}$ takes as input his identity $ID_{V}$, secret key 
$SK_{V}$, $\mathcal{I}$'s public key $PK_{I}$ and the public parameters
$PP$, and outputs  $(1,Tag_{V})$ if $ID_{V} \in J_{U}$ and the authentication tag
$Tag_{V}$ is valid; otherwise, he outputs $(0, Tag_{V})$ to indicate failure.
\medskip
\item{\sf Ticket-Trace}\,%
$(SK_{CV},PK_{CV},T_{U},PP)\rightarrow(ID_{U},J_{U})$. $\mathcal{CV}$ takes as inputs his
secret-public key pair $(SK_{CV},PK_{CV})$, a ticket $T_{U}$ and the public parameters $PP$, and outputs $\mathcal{U}$'s  identity $ID_{U}$
and $\mathcal{U}$'s  whole service set $J_{U}$. 
\medskip
\item{\sf Proxy-Key-Generation}%
$(\mathcal{V}'(ID_{V'})\leftrightarrow\mathcal{CA}(ID_{V},ID_{V'},$
$MSK,TP,PP)\rightarrow(RK_{\mathcal{V}\rightarrow\mathcal{V'}},\bot)$. This is
an interactive algorithm and is executed between a proxy verifier $\mathcal{V}'$
and $\mathcal{CA}$. $\mathcal{V}'$ takes as input its identity $ID_{V'}$ and
outputs a re-key $RK_{\mathcal{V}\rightarrow\mathcal{V}'}$ which enables
$\mathcal{V}'$ to validate tags on behalf of the verifier $\mathcal{V}$ in  the
time period $TP$. $\mathcal{CA}$ takes as input the identities of $\mathcal{V}$
ad $\mathcal{V}'$, the master secret key $MSK$ and the public parameters $PP$,
and output $\bot$. 
\medskip
\item{\sf Proxy-Ticket-Validation}\,%
$(\mathcal{U}(SK_{U},PK_{U},Tag_{V},PP) $ $\leftrightarrow 
\mathcal{V}'(SK_{V'},RK_{\mathcal{V}\rightarrow\mathcal{V}'},PP) \rightarrow
(\bot,(1,Tag_{V})/(0,Tag_{V}))$. This is an interactive algorithm and is
executed between a user $\mathcal{U}$ and a proxy verifier $\mathcal{V}'$.
$\mathcal{U}$ takes as input her secret-public key pair $(SK_{U},PK_{U})$,  the corresponding tag  $Tag_{V}$ and the public parameters $PP$,
and outputs $\bot$. $\mathcal{V}'$ takes as input its secret key $SK_{V'}$, the
re-key $RK_{\mathcal{V}\rightarrow\mathcal{V}'}$ and the public parameters $PP$,
and outputs $(1, Tag_{V})$ if $ID_{V}\in J_{U}$ and $Tag_{V}$ is valid;
otherwise, it outputs $(0,Tag_{V})$ to indicate failure.
\end{enumerate}

\begin{definition}
An anonymous Single Sign-On with proxy re-verification scheme  is correct if 

\begin{equation*}
\Pr\left[\begin{array}{l|l}
                         & {\sf Setup}(1^{\ell})\rightarrow 
                         \left(MSK,PP\right); \\
                        &  {\mbox{\em Ticket-Issuer-Reg}} (\mathcal{I}(ID_{I},SK_{I},\\
  \mbox{{\sf Ticket-}}       & PK_{I},PP)\leftrightarrow \mathcal{CA}(MSK,\\
  \mbox{ {\sf Validation}}                   &         PP))\rightarrow (\sigma_{I},(ID_{I},PK_{I}));\\
 (\mathcal{U}(SK_{U}, &  {\mbox{\em Ticket-Verifier-Reg}} (\mathcal{V}(ID_{V})\\
 PK_{U},  & \leftrightarrow    \mathcal{CA}(MSK, PP))\rightarrow\\
  Tag_{V},& ((\sigma_{V},SK_{V})(ID_{V}, SK_{V}));\\
 PP)\leftrightarrow    &  {\mbox{\em User-Reg}} (\mathcal{U}(ID_{U},SK_{U},PK_{U},\\
  \mathcal{V}((ID_{V}, & PP)\leftrightarrow \mathcal{CA}(MSK, PP))\\
 SK_{V},    & \rightarrow (\sigma_{U},(ID_{U},PK_{U}));\\
 PK_{I},            & {\mbox{\em Central-Verifier-Reg}}  (\mathcal{CV}(\\
 PP))             &ID_{CV},SK_{CV},PK_{CV},PP) \\
  \rightarrow(\bot, (1,                   & \leftrightarrow  \mathcal{CA}(MSK,PP)\rightarrow\\
  Tag_{V})  )                 &  (\sigma_{CV},(ID_{CV},PK_{CV}));\\
                  & {\mbox{\sf Ticket-Issuing}}(\mathcal{U}(SK_{U},PK_{U},\\
                   &J_{U}, \sigma_{U},PP)\leftrightarrow \mathcal{S}(SK_{I},PK_{I},\\
                   & PP))\rightarrow  (T_{U},J_{U});\\
                   & ID_{V}\in J_{U}
\end{array}
\right]=1,
\end{equation*}

\begin{equation*}
\Pr\left[\begin{array}{l|l}
                         & {\sf Setup}(1^{\ell})\rightarrow 
                         \left(MSK,PP\right); \\
                         &  {\mbox{\em Ticket-Issuer-Reg}} (\mathcal{I}(ID_{I},SK_{I},\\
                         &PK_{I},PP)\leftrightarrow  \mathcal{CA}(MSK,\\
                        &        PP))\rightarrow (\sigma_{S},(ID_{I},PK_{I}));\\
     & {\mbox{\em Ticket-Verifier-Reg}} (\mathcal{V}(ID_{V})\\
 {\sf Ticket-}    &\leftrightarrow \mathcal{CA}(MSK, PP))\rightarrow\\
{\sf Trace}  &  (\sigma_{V},(ID_{V},PK_{V}));\\
(  SK_{CV},  & {\mbox{\em User-Reg}} (\mathcal{U}(ID_{U},SK_{U}, PK_{U},\\
PK_{CV}, &PP)\leftrightarrow \mathcal{CA}(MSK,  PP))\\
T_{U},  PP))    &\rightarrow (\sigma_{U},(ID_{U},PK_{U}));\\
 \rightarrow (ID_{U},  & {\mbox{\em Central-Verifier-Reg}} (\mathcal{CV}(\\
 J_{U})& ID_{CV},SK_{CV},PK_{CV},PP)\\
 &\leftrightarrow \mathcal{CA}(MSK,PP))\rightarrow\\
& (\sigma_{CV},(ID_{CV},PK_{CV}));\\
              & {\mbox{\sf Ticket-Issuing}}(\mathcal{U}(SK_{U},PK_{U},\\
              & J_{U}, \sigma_{U},PP)\leftrightarrow \mathcal{I} 
              (SK_{I},PK_{I},\\
          &PP))\rightarrow  (T_{U},J_{U})
\end{array}
\right]=1
\end{equation*}

and

\begin{equation*}
\Pr\left[\begin{array}{l|l}
                         & {\sf Setup}(1^{\ell})\rightarrow 
                         \left(MSK,PP\right); \\
                         &  {\mbox{\em Ticket-Issuer-Reg}} (\mathcal{I}(ID_{I},SK_{I},\\
{\sf Proxy-}       & PK_{I},PP)\leftrightarrow  \mathcal{CA}(MSK,\\
 {\sf Ticket-}      &  PP))\rightarrow (\sigma_{I},(ID_{I},PK_{I}));\\
{\sf Validation}  &  {\mbox{\em Ticket-Verifier-Reg}} (\mathcal{V}(ID_{V})\\
(\mathcal{U}(SK_{U},           & \leftrightarrow   \mathcal{CA}(MSK,  PP))\rightarrow \\
  PK_{U}, & ((\sigma_{V},SK_{V}),(ID_{V},SK_{V}));\\
Tag_{V},                    &  {\mbox{\em User-Reg}} (\mathcal{U}(ID_{U},SK_{U},PK_{U},\\
PP)\leftrightarrow  & PP)\leftrightarrow \mathcal{CA}(MSK,PP))\\
 \mathcal{V}'(SK_{V'},         &\rightarrow (\sigma_{U},(ID_{U},PK_{U}));\\
RK_{\mathcal{V}\rightarrow\mathcal{V}'},      & {\mbox{\em Central-Verifier-Reg}} (\mathcal{CV}(\\
PK_{I}, PP))      &ID_{CV},SK_{CV},PK_{CV},PP)\\
  \rightarrow(\bot,    (1,       &\leftrightarrow \mathcal{CA}(MSK,PP))\rightarrow  \\
Tag_{V}) )    &(\sigma_{CV},(ID_{CV},PK_{CV}));\\
                & {\mbox{\sf Ticket-Issuing}}(\mathcal{U}(SK_{U},PK_{U},\\
                   & J_{U},\sigma_{U},PP)\leftrightarrow \mathcal{I}(SK_{I},PK_{I},\\
                   & PP))\rightarrow  (T_{U},J_{U});\\
                   & \mbox{{\sf Re-Key-Generation}}(\mathcal{V}'(ID_{V'})\\
                   &\leftrightarrow \mathcal{CA}(ID_{V},ID_{V'},MSK,TP,\\
                   &PP)\rightarrow(RK_{\mathcal{V}\rightarrow\mathcal{V'}},\bot);\\
                   & ID_{V}\in J_{U}
\end{array}
\right]=1.
\end{equation*}
\end{definition}

\subsection{Security Requirements}%
\label{sec-mod}%
The security model of our scheme  is defined by the following three games.

{\bf Unforgeability.} This is used to define the unforgeability of tickets, namely even if
users, verifiers and the central verifier collude, they cannot forge a valid
ticket. This game is formalised as follows.
\medskip

\noindent{\sf Setup.} %
$\mathcal{C}$ runs {\sf Setup}$(1^{\ell})\rightarrow (MSK,PP)$ and sends
$PP$ to $\mathcal{A}$. %
\medskip

\noindent{\sf Registration Query.} %
$\mathcal{A}$ can adaptively make the following queries.

\begin{enumerate}
\item{\em Ticket Issuer Registration Query.}  %
$\mathcal{C}$ runs $\mathcal{KG}(1^{\ell})\rightarrow(SK_{I},PK_{I})$ and
{\em Ticket-Seller-Reg} $(\mathcal{I}(ID_{I},SK_{I},PK_{I},PP)\leftrightarrow
\mathcal{CA}(MSK,PP))$ $\rightarrow(\sigma_{I},(ID_{I}, $ $PK_{I}))$, and
sends $(PK_{I},\sigma_{I})$ to $\mathcal{A}$. %

\item{\em Ticket Verifier Registration Query.} %
$\mathcal{A}$ submits an identity $ID_{V}$. 
$\mathcal{A}$ and
$\mathcal{C}$ run {\em Ticket-Verifier-Reg} $(\mathcal{V}(ID_{V})\leftrightarrow
\mathcal{CA}(MSK,PP))\rightarrow((\sigma_{V},SK_{V}),(ID_{V},SK_{V}))$.
$\mathcal{C}$ returns $(\sigma_{V},SK_{V})$ to $\mathcal{V}$;%

\item{\em User Registration Query.}  $\mathcal{A}$ submits an identity $ID_{U}$
and  and the corresponding public key $PK_{U}$.   $\mathcal{A}$ and
$\mathcal{C}$  run {\em User-Reg}$(\mathcal{U}(ID_{U},SK_{U},$
$PK_{U},PP)\leftrightarrow
\mathcal{CA}(MSK,$ $PP))\rightarrow(\sigma_{U},(ID_{U},PK_{U}))$.
$\mathcal{C}$ returns $\sigma_{U}$. 

\item{\em Central Verifier Registration Query.} $\mathcal{A}$ submits a central
verifier's identity $ID_{CV}$ and the corresponding public key $PK_{CV}$. $\mathcal{A}$ and $\mathcal{C}$  run  {\em
Central-Verifier-Reg}
$(\mathcal{CV}(ID_{CV},SK_{CV},PK_{CV},PP)\leftrightarrow \mathcal{CA}(MSK,$
$PP))\rightarrow(\sigma_{CV},(ID_{CV},PK_{CV}))$. $\mathcal{C}$ sends
$\sigma_{CV}$ to $\mathcal{A}$.
\end{enumerate}

\noindent{\sf Ticket Issuing Query.}  $\mathcal{A}$ adaptively submits  a set of
service information $J_{U}$. $\mathcal{C}$ runs {\sf Ticket-Issuing}
$\big(\mathcal{U}$ $(SK_{U},PK_{U},J_{U},\sigma_{U},PP)\leftrightarrow
\mathcal{I}(SK_{I},PK_{I},PP)\big)\rightarrow (T_{U}, J_{U})$ and
sends $T_{U}$ to $\mathcal{A}$. Let $TQ$ be an initially empty set 
consisting of the ticket
information queried by $\mathcal{A}$. $\mathcal{C}$ 
adds
$(T_{U},J_{U})$ into $TQ$. \medskip

\noindent{\sf Output.} $\mathcal{A}$ outputs a ticket
$T_{U^{*}}=\{Tag_{V^{*}}|V^{*} \in J_{U^{*}}\}\cup \{Tag_{CV}\}$ for a user
$\mathcal{U}^{*}$  with a set of service information $J_{U^{*}}$. $\mathcal{A}$
wins the game if {\sf Ticket-Validating}
$(\mathcal{U}(SK_{U^{*}},PK_{U^{*}},Tag_{V^{*}},PP) \leftrightarrow
\mathcal{V}((SK_{V^{*}},$ $PK_{V^{*}}),PK_{I},$
$PP))\rightarrow(\bot,(1,Tag_{V^{*}}))$ for all ${ID_{V}^{*}\in J_{U^{*}}}$ and
$(T_{U^{*}},J_{U^{*}})\notin TQ$.

\begin{definition}%
 An anonymous Single-Sign-On with proxy re-verification scheme  is
$(\varrho,\epsilon(\ell))$ unforgeable if all 
probabilistic
polynomial-time (PPT) adversaries $\mathcal{A}$ who makes $\varrho$  ticket
issuing queries can only win the above game with a negligible advantage, namely
\begin{equation*}%
\begin{split}
Adv_{\mathcal{A}}=\Pr\left[\begin{array}{cc}{\mbox{\sf 
Ticket-Validating}}(\mathcal{U}(SK_{U^{*}},PK_{U^{*}},\\ Tag_{V^{*}},PP)\leftrightarrow
 \mathcal{V}((SK_{V^{*}},PK_{V^{*}}),\\
PK_{I},PP))
\rightarrow(\bot,(1,Tag_{V^{*}}))
\end{array}\right]\leq \epsilon(\ell)
\end{split}
\end{equation*}
for all $ID_{V}^{*}\in J_{U^{*}}$.
\end{definition}
\medskip

{\bf Unlinkability.} This  is used to define the unlinkability,  namely even if some ticket verifiers
collude with  potential users, they cannot profile the whole service information
of other users. We assume that $\mathcal{I}$ and $\mathcal{CV}$ cannot be
compromised because they can know a user's whole service information by
themselves. The game is formalised  as follows.
\medskip

\noindent{\sf Setup.} $\mathcal{C}$ runs {\sf Setup}$(1^{\ell})\rightarrow 
(MSK,PP)$ and sends $PP$ to $\mathcal{A}$.
\medskip

\noindent{\bf Phase 1.} $\mathcal{A}$ can adaptively make the following queries. 
\medskip

\noindent{\sf Registration Query.} $\mathcal{A}$ adaptively makes the following registration queries.

\begin{enumerate}
\item{\em Ticket Issuer Registration Query.}  %
$\mathcal{C}$ runs $\mathcal{KG}(1^{\ell})\rightarrow(SK_{I},PK_{I})$ and {\em 
Ticket-Issuer-Registration} $(\mathcal{I}(ID_{I},SK_{I},$
$PK_{I},PP)\leftrightarrow
\mathcal{CA}(MSK,PP))\rightarrow(\sigma_{I},(ID_{I}, PK_{I}))$, and sends
$(ID_{I},PK_{I},$ $\sigma_{I})$ to $\mathcal{A}$.

\item{\em Ticket Verifier Registration Query.} %
Let $Corrupt_{V}$ be the set consisting of the identities of ticket verifiers
corrupted by $\mathcal{A}$. $\mathcal{A}$  submits a verifier's identity
$ID_{V}$: (1) if $ID_{V}\in Corrupt_{V}$,  $\mathcal{C}$ runs {\em 
Ticket-Verifier-Reg}$(\mathcal{V}(ID_{V})\leftrightarrow\mathcal{CA}(MSK,PP))
\rightarrow((\sigma_{V},SK_{V}),(ID_{V}, SK_{V}))$, and sends
$(\sigma_{V},SK_{V})$ to $\mathcal{A}$; (2) if $ID_{V}\notin Corrupt_{V}$,
$\mathcal{C}$ runs {\em Ticket-Verifier-Reg}
$(\mathcal{V}(ID_{V})\leftrightarrow\mathcal{CA}(MSK,PP))$
$\rightarrow((\sigma_{V},SK_{V}),(ID_{V},$ $ SK_{V}))$, and sends $\sigma_{V}$
to $\mathcal{A}$. Let $VK$ be a set consisting of the ticket verifier
registration query information $(ID_{V},SK_{V},\sigma_{V})$.

\item{\em User Registration Query.} $\mathcal{A}$  submits a user's identity
$ID_{U}$ and runs $\mathcal{KG}(1^{\ell})\rightarrow(SK_{U},PK_{U})$.
$\mathcal{A}$ and $\mathcal{C}$ run  {\em 
User-Registration}$(\mathcal{U}(ID_{U},SK_{U},PK_{U},PP)\leftrightarrow
\mathcal{CA}(MSK,PP))$ $\rightarrow(\sigma_{U},(ID_{U},PK_{U}))$.
$\mathcal{C}$ sends  $\sigma_{U}$ to $\mathcal{A}$. %

\item{\em Central Verifier Registration Query.}  $\mathcal{C}$ runs 
$\mathcal{KG}(1^{\sw{\ell}})\rightarrow (SK_{CV},PK_{CV})$ and  {\em  
Central-Verifier-Reg} $(\mathcal{CV}(ID_{CV},SK_{CV},PK_{CV},PP)\leftrightarrow 
\mathcal{CA}(MSK,PP))\rightarrow(\sigma_{CV},(ID_{CV},PK_{CV}))$. 
$\mathcal{C}$ returns $(PK_{CV},$ $\sigma_{CV})$ to $\mathcal{A}$.  
\end{enumerate}

\noindent{\sf Ticket Issuing Query.}  %
$\mathcal{A}$  adaptively submits  a set of {service} information $J_{U}$ to
$\mathcal{C}$. $\mathcal{C}$ runs {\sf Ticket-Issuing}$(\mathcal{U}(SK_{U},$
$PK_{U},J_{U},\sigma_{U},PP)\leftrightarrow
\mathcal{I}(SK_{I},PK_{I},\sigma_{I},PP))\rightarrow (T_{U},J_{U})$. Let $TQ$ be
 an  initially empty set which consists of the ticket information queried by
$\mathcal{A}$. $\mathcal{C}$ adds $(T_{U},J_{U})$ into $TQ$ and sends $T_{U}$ to
$\mathcal{A}$. 
\medskip

\noindent{\sf Ticket Trace Query.} %
$\mathcal{A}$  adaptively submits a ticket $T_{U}$. $\mathcal{C}$  runs {\sf
Ticket-Trace}$(SK_{CV},
PK_{CV},T_{U},PP)\rightarrow(ID_{U},J_{U})$, and returns $(ID_{U},J_{U})$ to $\mathcal{A}$
if $T_{U}\in TQ$. Let $TTQ$ be  an  initially empty set 
which consists of the 
ticket trace information queried
by $\mathcal{A}$. $\mathcal{C}$ adds $(T_{U},ID_{U},J_{U})$ 
into
$TTQ$. %
\medskip

\noindent{\sf Proxy Key Generation Query.} %
$\mathcal{A}$ adaptively submits two identities $ID_{V}$ and $ID_{V'}$ and
$\mathcal{C}$ runs {\sf
Pxoy-Key-Generation}$(\mathcal{V}'(ID_{V'})\leftrightarrow\mathcal{CA}(ID_{V},ID_{V'},MSK,TP,PP)$
 $\rightarrow(RK_{\mathcal{V}\rightarrow\mathcal{V'}},\bot)$ and sends 
$RK_{\mathcal{V}\rightarrow\mathcal{V}'}$ to $\mathcal{A}$. Let $PQ$ be the 
initially empty set consisting of the proxy key generation query. 
$\mathcal{C}$ adds 
$(\mathcal{V},\mathcal{V}',TT,RK_{\mathcal{V}\rightarrow\mathcal{V}'})$ into 
$PQ$. \medskip

\noindent{\sf Proxy Ticket Validation Query.} %
$\mathcal{A}$ adaptively submits $(T_{U},ID_{V},ID_{V'})$. $\mathcal{C}$ runs
{\sf Proxy-Key-Generation}$(\mathcal{V}'(ID_{V'})
\leftrightarrow\mathcal{CA}(ID_{V},ID_{V'},MSK,TP,PP)$
$\rightarrow(RK_{\mathcal{V}\rightarrow\mathcal{V'}},\bot)$ and {\sf
Proxy-Ticket-Validation}$(\mathcal{U}(T_{U},PP)
\leftrightarrow\mathcal{V}'(SK_{V'},$
$RK_{\mathcal{V}\rightarrow\mathcal{V}'},PP)
\rightarrow(\bot,(1,Tag_{V})/(0,Tag_{V}))$ with $\mathcal{A}$. $\mathcal{C}$
sends $(1,Tag_{V})$ to $\mathcal{A}$ if $ID_{V}\in J_{U}$ and $Tag_{V}$ is
valid; otherwise, $(0,Tag_{V})$ is sent to $\mathcal{A}$ to indicate a failure.
\medskip

\noindent{\bf Challenge.}  %
$\mathcal{A}$ submits two verifiers $V_{0}^{*}$ and $V_{1}^{*}$ with the
limitation that $ID_{V_{0}^{*}},ID_{V_{1}^{*}}\notin Corrupt_{V}$ and 
$(\mathcal{V}_{0}^{*},\mathcal{V}_{1}^{*},TP,RK_{\mathcal{V}_{0}^{*}\rightarrow\mathcal{V}_{1}^{*}}), (\mathcal{V}_{0}^{*},\mathcal{V}_{1}^{*},TP,RK_{\mathcal{V}_{1}^{*}\rightarrow\mathcal{V}_{0}^{*}})\notin
PQ$. $\mathcal{C}$ flips an unbiased coin with $\{0,1\}$ and obtains a bit
$b\in\{0,1\}$.  $\mathcal{C}$ sets $J_{U^{*}}=\{V_{b}^{*}\}$ and runs {\sf
Ticket-Issuing}$(\mathcal{U}(SK_{U^{*}},PK_{U^{*}},J_{U^{*}},$
$\sigma_{U^{*}},PP)\leftrightarrow
\mathcal{I}(SK_{I},PK_{I},PP))\rightarrow (T_{U^{*}},J_{U^{*}})$
where $T_{U^{*}}=(Tag_{V_{b}}^{*},Tag_{CV})$ and $Tag_{V_{b}}^{*}\notin T_{U}$
for all $(T_{U},J_{U})\in TQ$.  $\mathcal{C}$ sends $T_{U^{*}}$ to
$\mathcal{A}$. %
\medskip

\noindent{\bf Phase 2.} It is the same as in {\bf Phase 1}, except  with the
limitation that $ID_{V_{0}^{*}},ID_{V_{1}^{*}}\notin Corrupt_{V}$ and 
$(\mathcal{V}_{0}^{*},\mathcal{V}_{1}^{*},TP,RK_{\mathcal{V}_{0}^{*}\rightarrow\mathcal{V}_{1}^{*}}), (\mathcal{V}_{0}^{*},\mathcal{V}_{1}^{*},TP,RK_{\mathcal{V}_{1}^{*}\rightarrow\mathcal{V}_{0}^{*}})\notin
PQ$.
\medskip

\noindent{\bf Output.} $\mathcal{A}$ outputs his guess $b'$ on $b$.
$\mathcal{A}$ wins the game if $b'=b$.

\begin{definition}
An anonymous Single-Sign-On with proxy re-verification scheme is 
$(\varrho_{1},\varrho_{2},\varrho_{3},\epsilon(\ell))$ unlinkable 
if  all probabilistic polynomial-time  (PPT) adversaries $\mathcal{A}$ making 
at most $\varrho_{1}$ ticket issuing queries, $\varrho_{2}$ ticket trace queries and $\varrho_{3}$ proxy key 
generation queries  can win the above game with a negligible advantage, namely
\begin{equation*}
Adv_{\mathcal{A}}=\left|\Pr\left[b'=b\right]-\frac{1}{2}\right|\leq \epsilon(\ell).
\end{equation*}
\end{definition}

\medskip
{\bf Traceability.} This is used to formalise the traceability of tickets, namely even if a
group of users  collude, they cannot generate a ticket which $\mathcal{CV}$ would not catch as belonging to some member of the colluding
group. We suppose that the ticket issuer is honest. This game is formalised as follows.
\medskip

\medskip

\noindent{\sf Setup.} $\mathcal{C}$ runs {\sf Setup}$(1^{\ell})\rightarrow
(MSK,PP)$ and sends $PP$ to $\mathcal{A}$. \medskip

\noindent{\sf Registration Query.} $\mathcal{A}$ can  adaptively make the following queries.

\begin{enumerate}
\item{\em Ticket Issuer Registration Query.}  %
$\mathcal{C}$ runs
$\mathcal{KG}(1^{\ell})\rightarrow(SK_{I},PK_{I})$ and {\em 
Ticket-Issuer-Reg} $(\mathcal{I}(ID_{I},SK_{I},$ $PK_{I},PP)\leftrightarrow
\mathcal{CA}(MSK,PP))\rightarrow(\sigma_{I},(ID_{I}, $ $PK_{I}))$,
and sends $(PK_{I},\sigma_{I})$ to $\mathcal{A}$. 

\item{\em  Ticket
Verifier Registration Query.} $\mathcal{A}$ submits an identity $ID_{V}$ and
runs {\em Ticket-Verifier-Registration}
$(\mathcal{V}(ID_{V},SK_{V},PK_{V},PP)\leftrightarrow
\mathcal{CA}(MSK,PP))\rightarrow(\sigma_{V},(ID_{V},$ $PK_{V}))$ with
$\mathcal{C}$. $\mathcal{C}$ sends $(SK_{V},\sigma_{V})$ to $\mathcal{A}$. %

\item{\em User Registration Query.}  %
$\mathcal{A}$ submits an identity $ID_{U}$ and runs
$\mathcal{KG}(1^{\ell})\rightarrow(SK_{U},PK_{U})$.  $\mathcal{A}$ and
$\mathcal{C}$ run {\em	User-Reg}$(\mathcal{U}(ID_{U},SK_{U},$
$PK_{U},PP)\leftrightarrow
\mathcal{CA}(MSK,PP))\rightarrow(\sigma_{U},(ID_{U},PK_{U}))$.
$\mathcal{C}$ sends $\sigma_{U}$ to $\mathcal{A}$.  Let $QK_{U}$ be an 
initially empty set
which consists of the users' identities selected by $\mathcal{A}$ to make
registration query. %

\item{\em Central Verifier Registration Query.} %
$\mathcal{C}$  runs $\mathcal{KG}(1^{\ell})\rightarrow(SK_{CV},PK_{CV})$
and {\em Central-Verifier-Reg}
$(\mathcal{CV}(ID_{CV},SK_{CV},PK_{CV},PP)\leftrightarrow \mathcal{CA}(MSK,$
$PP))\rightarrow(\sigma_{CV},(ID_{CV},PK_{CV}))$.  $\mathcal{C}$ sends
$(PK_{CV},\sigma_{CV})$ to $\mathcal{A}$.
\end{enumerate}

\noindent{\sf Ticket Issuing Query.}  %
$\mathcal{A}$ adaptively submits a set of service information $J_{U}$.
$\mathcal{C}$ and $\mathcal{A}$ runs {\sf Ticket-Issuing} $\big(\mathcal{U}$
$(SK_{U},$ $PK_{U},J_{U},\sigma_{U},PP)\leftrightarrow
\mathcal{I}(SK_{I},PK_{I},PP)\big)\rightarrow (T_{U}, J_{U})$, and
sends $T_{U}$ to $\mathcal{A}$. Let $TQ$ be an initially empty set which 
consists of the ticket
information queried by $\mathcal{A}$. $\mathcal{C}$ 
adds
$(T_{U},J_{U})$ into $TQ$. %
\medskip

\noindent{\bf Output.} $\mathcal{A}$ outputs a ticket
$T_{U^{*}}=\{Tag_{V^{*}}|V^{*}\in J_{U^{*}}\}\cup \{Tag_{CV}^{*}\}$ for a user
$\mathcal{U}^{*}$  with a set of service information $J_{U^{*}}$. $\mathcal{A}$
wins the game if {\sf Ticket-Trace}
$(SK_{CV},PK_{CV},Tag_{CV}^{*},T_{U}^{*},PP)\rightarrow
(ID_{\tilde{U}},J_{\tilde{U}}))$ with $ID_{\tilde{U}}\notin QK_{U} $
or $ID_{\tilde{U}}\neq ID_{U^{*}}$.

\begin{definition}
An anonymous Single-Sign-On with proxy re-verification scheme is
$(\varrho,\epsilon(\ell))$ traceable if  all probabilistic
polynomial-time adversaries $\mathcal{A}$ who makes $\varrho$  ticket
issuing queries can win the above game with a negligible advantage, namely
\begin{equation*}%
\begin{split}
Adv_{\mathcal{A}}&=\Pr\left[\begin{array}{c|c}&\mbox{{\sf Ticket-Trace}}\\
ID_{\tilde{U}}\notin QK_{U} 
~\mbox{or}& (SK_{CV},PK_{CV},\\
~ ID_{U^{*}}\neq ID_{\tilde{U}}\in QK_{U} 
&Tag_{CV}^{*},T_{U}^{*},PP)\\
& \rightarrow (ID_{\tilde{U}},J_{\tilde{U}})
\end{array}\right]\\
 & \leq \epsilon(\ell).
 \end{split}
\end{equation*}
\end{definition}

\section{Preliminaries}\label{preli}
In this section, preliminaries used in this paper are introduced.
\subsection{Bilinear Group}

Let $\mathbb{G}_{1}$, $\mathbb{G}_{2}$ and $\mathbb{G}_{\tau}$ be cyclic groups
with prime order $p$.  A map
$e:\mathbb{G}_{1}\times\mathbb{G}_{2}\rightarrow\mathbb{G}_{\tau}$ is a bilinear
map/pairing if it satisfies the following properties \cite{bf:ibe2001}:
{\sf (1) Bilinearity:} For all $g\in\mathbb{G}_{1}$, $h\in\mathbb{G}_{2}$ and 
$x,y\in\mathbb{Z}_{p}$, $e(g^{x},h^{y})=e(g^{y},h^{x})=e(g,h)^{xy}$;
{\sf  (2) Non-degeneration:} For all $g\in\mathbb{G}_{1}$ and 
$h\in\mathbb{G}_{2}$, $e(g,h)\neq 1_{\tau}$ where $1_{\tau}$ is the identity 
element in $\mathbb{G}_{\tau}$;
{\sf (3) Computability:} For all $g\in\mathbb{G}_{1}$ and 
$h\in\mathbb{G}_{2}$, there exists an efficient algorithm to compute $e(g,h)$.

Let $\mathcal{BG}(1^{\ell})
\rightarrow(e,p,\mathbb{G}_{1},\mathbb{G}_{2},\mathbb{G}_{\tau})$ be a bilinear
group generator which takes as input a security parameter $1^{\ell}$ and outputs
a bilinear group $(e,p,\mathbb{G}_{1},\mathbb{G}_{2},\mathbb{G}_{\tau})$.
Bilinear maps can be divided  into three types \cite{gps:pairing2008}: Type-I:
$\mathbb{G}_{1}=\mathbb{G}_{2}$; Type-II: $\mathbb{G}_{1}\neq \mathbb{G}_{2}$
but there exist an efficient map:
$\phi:\mathbb{G}_{2}\rightarrow\mathbb{G}_{1}$; Type-III: $\mathbb{G}_{1}\neq
\mathbb{G}_{2}$ but there is no efficient map between $\mathbb{G}_{1}$ and
$\mathbb{G}_{2}$. Type-III pairings are the most efficient pairings \cite{cm:dabdh2011}. 
Our scheme is based on Type-III pairings where the size
of elements in $\mathbb{G}_{1}$ is short (160 bits).

\subsection{Complexity Assumptions}
\begin{definition}{\sf (Discrete Logarithm (DL) Assumption \cite{g:dl1993})}
Let $\mathbb{G}$ be a cyclic group with prime order $p$ and $g$ be a generator of $\mathbb{G}$. Given $Y\in \mathbb{G}$, we say that the DL assumption holds on $\mathbb{G}$ if all PPT adversaries can output a number $x\in\mathbb{Z}_{p}$ such that $Y=g^{x}$ with a negligible advantage, namely 
%
$Adv_{\mathcal{A}}^{DL}=\Pr\left[Y=g^{x}|\mathcal{A}(p,g,\mathbb{G},Y)\rightarrow
 x\right]\leq \epsilon(\ell).$
\end{definition}
The proof of the traceability property of our scheme is reduced to the $DL$ assumption.

\begin{definition}{\sf (Decisional  Bilinear Diffie-Hellman (DBDH) Assumption \cite{bf:ibe2001})}
Let $\mathcal{BG}(1^{\ell})\rightarrow(e,p,\mathbb{G},\mathbb{G}_{\tau})$ 
where $\mathbb{G}_{1}=\mathbb{G}_{2}=\mathbb{G}$ and $g$ be a generator of 
$\mathbb{G}$. Suppose that 
$a,b,c\stackrel{R}{\leftarrow}\mathbb{Z}_{p}$. 
Given a tuple $\mathbb{T}=(g,g^{a},g^{b},g^{c}, \Upsilon)$, we say that the 
DBDH assumption holds on $(e,p,\mathbb{G},\mathbb{G}_{\tau})$  if all  
PPT adversary $\mathcal{A}$ can distinguish 
$\Upsilon=e(g,g)^{abc}$ from a random element $R\in\mathbb{G}_{\tau}$ with a 
negligible advantage, namely 
$Adv_{\mathcal{A}}^{DBDH}=\Big|\Pr[\mathcal{A}(\mathbb{T},\Upsilon=e(g,g)^{abc})=1]-\Pr[\mathcal{A}(\mathbb{T},\Upsilon=R)=1]\Big|
\leq \epsilon(\ell).$
\end{definition}
 The security of the Boneh-Franklin IBE used to implement flexible verification was reduced to the DBDH assumption.

\begin{definition}{\sf (Decisional asymmetric Bilinear Diffie-Hellman (DaBDH) Assumption \cite{cm:dabdh2011})}
Let $\mathcal{BG}(1^{\ell})
\rightarrow(e,p,\mathbb{G}_{1},\mathbb{G}_{2},\mathbb{G}_{\tau})$ and $g$,
$\mathfrak{g}$ be generators of $\mathbb{G}_{1}$ and $\mathbb{G}_{2}$,
respectively. Suppose that $a,b,c\stackrel{R}{\leftarrow}\mathbb{Z}_{p}$.
Given a tuple
$\mathbb{T}=(g,\mathfrak{g},g^{a},g^{b},g^{c},\mathfrak{g}^{b},\mathfrak{g}^{c},
\Upsilon)$, we say that the DaBDH assumption holds on
$(e,p,\mathbb{G}_{1},\mathbb{G}_{2},\mathbb{G}_{\tau})$  if all PPT adversaries can distinguish $\Upsilon=e(g,\mathfrak{g})^{abc}$
from a random element $R\in\mathbb{G}_{\tau}$ with a negligible advantage, namely
$Adv_{\mathcal{A}}^{DaBDH}=\Big|\Pr[\mathcal{A}
(\mathbb{T},\Upsilon=e(g,\mathfrak{g})^{abc})=1]-\Pr[\mathcal{A}(\mathbb{T},\Upsilon=R)=1]\Big| \leq \epsilon(\ell).$%
\end{definition}%
The DaBDH assumption is used to prove the unlinkablity of our scheme.

\begin{definition}{\sf ((JoC Version) $q$-Strong Diffie-Hellman (JoC-$q$-SDH) 
Assumption \cite{bb:2008})} Let 
$\mathcal{BG}(1^{\ell})\rightarrow(e,p,\mathbb{G}_{1},\mathbb{G}_{2},$ 
$\mathbb{G}_{\tau})$. Given a $(q+3)$-tuple 
$(g,g^{x},\cdots,g^{x^{q}},\mathfrak{g},
\mathfrak{g}^{x})\in\mathbb{G}_{1}^{q+1}\times\mathbb{G}_{2}^{2}$, we say that 
the JoC-$q$-SDH  assumption holds on 
$(e,p,\mathbb{G}_{1},\mathbb{G}_{2},\mathbb{G}_{\tau})$ if all
PPT adversaries $\mathcal{A}$ can output 
$(c,g^{\frac{1}{x+c}})\in\mathbb{Z}_{p}\times\mathbb{G}_{1}$ with a negligible 
advantage, namely
$Adv_{\mathcal{A}}^{\mbox{JOC-q-SDH}}=
\Pr\left[(c,g^{\frac{1}{x+c}})\leftarrow\mathcal{A}(g,g^{x},\cdots,g^{x^{q}},
\mathfrak{g},\mathfrak{g}^{x})\right]
 \leq\epsilon(\ell),$
where $c\in\mathbb{Z}_{p}\setminus\{-x\}$.
\end{definition}
The unforgeability of our scheme is reduced to JoC-$q$-SDH assumption.

\subsection{Zero-Knowledge Proof}

We follow the definition  introduced by Camenish and Stadler in \cite{cs:1997}
 and formalised by Camenish {\em et al.} in
\cite{cky:2009}.
By 
$\mbox{PoK:}\{(x_{1},x_{2},x_{3}): 
\Upsilon=g^{x_{1}}h^{x_{2}}~\wedge~\tilde{\Upsilon}=\mathfrak{g}^{x_{1}} 
\mathfrak{h}^{x_{3}}\},$
we denote a zero knowledge proof on knowledge of integers $x_{1}$, $x_{2}$ and
$x_{3}$ such that $\Upsilon=g^{x_{1}}h^{x_{2}}$ and
$\tilde{\Upsilon}=\mathfrak{g}^{x_{1}} \mathfrak{h}^{x_{3}}$ hold on the groups
$\mathbb{G}=\langle g \rangle=\langle h \rangle$ and $\tilde{\mathbb{G}}=\langle
\mathfrak{g} \rangle=\langle \mathfrak{h} \rangle$, respectively.  The
convention is that the letters in the parenthesis $(x_{1},x_{2},x_{3})$ stand
for the knowledge which is being proven, while the other parameters are known by
the verifier.

\subsection{BBS+ Signature}

This signature was proposed by Au {\em et al.} \cite{asm:2006}.
Its security was reduced to the $q$-SDH assumption in  Type-II pairing setting in  \cite{asm:2006}. Recently, Camenisch {\em et al.} \cite{cdl:daa}  reduced its security to the JoC-$q$-SDH assumption
in  Type-III pairing setting.

\begin{theorem}(Camenisch\etal\cite{cdl:daa})%
\label{theo:BBS}%
The BBS+ signature is  existentially unforgeable against adaptive chosen message attacks (EU-CMA) if the JoC-$q$-SDH assumption holds on the bilinear group $(e,p,\mathbb{G}_{1},\mathbb{G}_{2},\mathbb{G}_{\tau})$.
\end{theorem}

\subsection{Boneh-Franklin Identity-Based Encryption}

Boneh and Franklin \cite{bf:ibe2001} proposed the first IBE scheme  based on the Type-I pairing: $e:\mathbb{G}\times\mathbb{G}\rightarrow\mathbb{G}_{\tau}$. 

\begin{theorem} (Boneh and Franklin \cite{bf:ibe2001}) %
\label{theo:IBE}%
This IBE scheme is secure against chosen-plaintext
attack (CPA) if the DBDH assumption holds on the bilinear map group 
$(e,p,\mathbb{G},\mathbb{G}_{\tau})$.

\end{theorem}

Abdalla {\em et al.} \cite{abckk:se2005} observed that Boneh-Franklin IBE \cite{bf:ibe2001} is an anonymous IBE scheme where ciphertext does not release the identity  of the receiver. Chatterjee and Menezes \cite{cm:dabdh2011} transferred Boneh-Franklin IBE scheme from Type-I pairing setting to Type-III pairing setting, and claimed that the security of the transferred scheme can be reduced to DaBDH assumption.
In this paper, the Boneh-Franklin \cite{bf:ibe2001} IBE scheme is applied 
to implement proxy re-verification.

\section{Construction of our scheme}%
\label{const}%

\subsection{Formal Construction} 
The formal construction of our ASSO with proxy re-verification scheme
including messages sent between its entities and their relevant
computations is presented in Fig. \ref{fig.setup}, Fig. \ref{fig.reg}, Fig.
\ref{fig.t-i}, Fig. \ref{fig.t-v}, Fig. \ref{fig.t-t}, Fig. \ref{fig.p-k-g} and
Fig. \ref{fig.p-t-v}. Notably, Fig. \ref{fig.p-k-g} and
Fig. \ref{fig.p-t-v} are  new in our scheme compared to Han\etal's construction 
in \cite{hcsts:asso2018} and Fig. \ref{fig.reg}, Fig.
\ref{fig.t-i} have been modified to reflect the IBPRE scheme used.

\subsection{High-Level Overview} 
At a high level, our scheme works as follows. 
\medskip

\noindent{\sf Setup.} %
$\mathcal{CA}$ initializes the systems and generates a master secret key
$MSK=(\alpha,\beta)$ and the corresponding public parameters $PP$. Actually,
$\alpha$ is used to issue credentials to $\mathcal{I}$, $\mathcal{U}$ and
$\mathcal{CV}$ when they join the system, while $\beta$ is used to issue secret
keys to $\mathcal{V}$s.
\medskip

\noindent{\sf Registration.} %
When joining the system, $\mathcal{I}$, $\mathcal{U}$ and $\mathcal{CV}$
generate their secret-public key pairs $(x_{i},Y_{I},\tilde{Y}_{I})$,
$(x_{u},Y_{U})$ and $(x_{cv},Y_{CV})$, and register with the $\mathcal{CA}$
by sending their identities $(ID_{I},ID_{U},ID_{CV})$ and public keys
$((Y_{I},\tilde{Y}_{I}),Y_{U},Y_{CV})$, respectively. Finally, $\mathcal{I}$,
$\mathcal{U}$ and $\mathcal{CV}$ obtain their credentials
$(d_{i},e_{i},\sigma_{I})$, $(d_{u},e_{u},\sigma_{U})$ and
$(d_{cv},e_{cv},\sigma_{CV})$ from $\mathcal{CA}$, respectively. Note,
$(d_{i},e_{i},\sigma_{I})$, $(d_{u},e_{u},\sigma_{U})$ and
$(d_{cv},e_{cv},\sigma_{CV})$ are generated by $\mathcal{CA}$ using the master
secret key $\alpha$ and are BBS+ signatures on the public keys $Y_{I}$, $Y_{U}$
and $Y_{CV}$, respectively. When joining the system, $\mathcal{V}$s only
submit  their identities to $\mathcal{CA}$. $\mathcal{CA}$ uses the master
secret key $\beta$ to generate a secret key $SK_{V}$ for the identity
$ID_{V}$ of $\mathcal{V}$. This is one of the main differences in the scheme's
construction compared to Han\etal\cite{hcsts:asso2018} where the verifiers
generate their own secret-public key pairs. Moving this generation to the
$\mathcal{CA}$ is required to facilitate the proxy re-verification. 
Furthermore, the $\mathcal{CA}$ generates a credential
$(d_{v},e_{v},\sigma_{V})$ for $\mathcal{V}$ which is a BBS+ signature on
$ID_{V}$. $\mathcal{CV}$ stores $((d_{v},e_{v},\sigma_{V}),SK_{V})$, and sends
them to $\mathcal{V}$.
\medskip

\noindent{\sf Ticket Issuing.} %
To buy a ticket, $\mathcal{U}$ determines her service information $J_{U}$
consisting of the identities of the corresponding $\mathcal{V}$ whose services
$\mathcal{U}$ wants to access. Furthermore, for each $ID_{V}\in J_{U}$,
$\mathcal{U}$ generates a pseudonym $(P_{V},Q_{V})$ using her secret key and
proves to $\mathcal{I}$ that she is a registered user and the pseudonyms are
generated correctly ($\prod_{U}^{1}$ ). If the proof is correct, for each
$ID_{V}\in J_{U}$, $\mathcal{I}$ generates an authentication tag
$Tag_{V}=((P_{V},Q_{V}),(E_{V}^{1},$
$E_{V}^{2},E_{V}^{3},K_{V},Text_{1},Text_{2}),(s_{v},w_{v},z_{v},Z_{V}))$.
Within the tag $(E_{V}^{1},E_{V}^{2})$ are used by $\mathcal{V}$ to validate
$Tag_{V}$ while $(E_{V}^{1},E_{V}^{2},E_{V}^{3})$ are used by a proxy verifier
$\mathcal{V}'$ to validate $Tag_{V}$ on behalf of $\mathcal{V}$ in a specified
time period TP. Since $TP$ is embedded in $E_{V}^{3}$ it is used to restrict the
time of proxy re-verification to reflect that time period. In a rail
application, a $TP$ could be the travel day printed on the ticket  (e.g.
September 1, 2018) and can be decided by the ticket issuer.

Additionally, within the tag  $((P_{V},Q_{V}),E_{V}^{2},K_{V})$ are used by
$\mathcal{CV}$ to de-anonymize $\mathcal{U}$'s identity and trace her service
requests. Note also that $s_{v}$ is  the serial number of $Tag_{V}$ and
$(w_{v},z_{v},Z_{V})$ is a BBS+ signature on $s_{v}$. To prevent $\mathcal{U}$
from combing the authentication tags in different tickets, $\mathcal{I}$
generates another BBS+ signature $(w,z,Z)$ on the ticket issue number
$s=H_{1}(s_{1}||s_{2}||\cdots||s_{|J_{U}|})$. The ticket is
$T_{U}=\{Tag_{V}|ID_{V}\in J_{U}\}\cup (s,w,z,Z)$.


\medskip

\noindent{\sf Ticket Validation.} %
When validating a ticket, $\mathcal{V}$ sends its identity $ID_{V}$ to
$\mathcal{U}$. $\mathcal{U}$ selects the corresponding tag $Tag_{V}$, and then
sends  it to $\mathcal{V}$ with a proof of the knowledge ($\prod_{U}^{2}$ ) of
the secrets included in the pseudonyms $(P_{V},Q_{V})$. $\mathcal{V}$ validates
the tag $Tag_{V}$ by checking the proof and the signature. However, in the case
that $\mathcal{U}$ needs to confirm whether $\mathcal{V}$ is a designated
verifier, $\mathcal{V}$  sends  $ID_{V}$ and his credential
$(d_{v},e_{v},\sigma_{V})$ to $\mathcal{U}$. Then, $\mathcal{U}$ checks
$e(\sigma_{V},Y_{A}\mathfrak{g}^{e_{v}})\stackrel{?}{=}e(g_{1}g_{2}^{d_{v}}
\tilde{g}^{H_{1}(ID_{V})})$. If it holds, $\mathcal{V}$ is a designated
verifier; otherwise,  $\mathcal{V}$ is not. In this paper, we assume that
$\mathcal{V}$ is clear and $\mathcal{U}$ does not need to confirm it. For
example in the rail scenario, the verifier/station is clear to $\mathcal{U}$.

\medskip
 
\noindent{\sf Ticket Trace.} %
To de-anonymize  a user and trace her service requests, $\mathcal{CV}$
initialises a set $\Omega_{U}$. Given a ticket $T_{U}$, $\mathcal{CV}$ uses his
secret key to de-anonymize $\mathcal{U}$ from the pseudonyms $(P_{V},Q_{V})$ for
$ID_{V}\in J_{U}$ and traces the service request from $(E_{V}^{2},K_{V})$.
Finally, $\mathcal{CV}$ can determine $\mathcal{U}$'s service requests by
recording all the identities $ID_{V}\in\Omega_{U}$.

\noindent{\sf Proxy Key Generation.} %
In the case that a verifier $\mathcal{V}$ is unavailable,
$\mathcal{CA}$ can authorize a proxy verifier $\mathcal{V}'$ to validate the tag
$Tag_{V}$ in a ticket $T_{U}$ by issuing a re-key
$RK_{\mathcal{V}\rightarrow\mathcal{V}'}$ to $\mathcal{V}'$.
$RK_{\mathcal{V}\rightarrow\mathcal{V}'}$ is generated by using both  secret
keys $SK_{V}$ and $SK_{V'}$. To limit the proxy verification period, a
time period $TP$, which is embedded in $E^3_V$ during the Ticket Issuing, is
also embedded in $RK_{\mathcal{V}\rightarrow\mathcal{V}'}$so that only
tickets within that $TP$ period can be validated by the proxy verifier. 
To prevent an unauthorised verifier from claiming to be a legal proxy, the
$\mathcal{CA}$ or another trusted third party should broadcast the proxy 
information $(ID_{V'})$ to both $\mathcal{U}$ and $\mathcal{V}'$. For example, 
in a rail scenario, when a station $\mathcal{V}$ is 
unavailable and an alternative plan is provided, both the user $\mathcal{U}$ 
and the proxy $\mathcal{V}'$ need to be notified.
\medskip

\noindent{\sf Proxy Ticket Validation.} %
To verify a tag $Tag_{V}$ on behalf of $\mathcal{V}$, $\mathcal{V}'$ sends the
identity $ID_{V}'$ to $\mathcal{U}$. $\mathcal{U}$ returns the tag $Tag_{V}$ to
$\mathcal{V}'$ and proves the knowledge included in $Tag_{V}$. If the proof is
correct, $\mathcal{V}'$ validates $Tag_{V}$ by using his secret key $SK_{V'}$
and the re-key $RK_{\mathcal{V}\leftarrow\mathcal{V}'}$. Both the user and the
proxy verifier $\mathcal{V}'$ know that $\mathcal{V}'$ is a proxy for the
verifier $\mathcal{V}$ as discussed above. For example, in a transport
application a public announcement would identify the alternative route and hence
the corresponding proxy verifier to the user.

\begin{figure*}[]
\centering
\fbox{
\begin{minipage}{17.5cm}
\begin{center}{\sf Setup}$(1^{\lambda})$\end{center}%

$\mathcal{CA}$ runs
$\mathcal{BG}(1^{\ell})\rightarrow(e,p,\mathbb{G}_{1},\mathbb{G}_{2},
\mathbb{G}_{\tau})$ with
$e:\mathbb{G}_{1}\times\mathbb{G}_{2}\rightarrow\mathbb{G}_{\tau}$.  Let
$\tilde{g},\bar{g},g_{1},g_{2},g_{3}$ be  generators of $\mathbb{G}_{1}$  and
$\mathfrak{g}$ be generators of $\mathbb{G}_{2}$. Suppose that
$H_{1}:\{0,1\}^{*}\rightarrow\mathbb{Z}_{p}$, $H_{2}:\{0,1\}\rightarrow
\mathbb{G}_{2}$ and $H_{3}:\{0,1\}^{*}\rightarrow \{0,1\}^{\ell'} ~
(\ell'\leq\ell)$ are cryptographic hash functions. $\mathcal{CA}$ selects
$\alpha,\beta,\stackrel{R}{\leftarrow}\mathbb{Z}_{p}$ and
$\vartheta_{1},\vartheta_{2}\stackrel{R}{\leftarrow}\mathbb{G}_{2}$.
$\mathcal{CA}$ computes ${Y}_{A}=\mathfrak{g}^{\alpha}$ and
$\tilde{Y}_{A}=\tilde{g}^{\beta}$. The master secret key is $MSK=(\alpha,\beta)$
and the public parameters are
$PP=(e,p,\mathbb{G}_{1},\mathbb{G}_{2},\mathbb{G}_{\tau},
\tilde{g},\bar{g},g_{1},g_{2},g_{3},\mathfrak{g},\vartheta_{1},
\vartheta_{2},Y_{A},\tilde{Y}_{A},H_{1},H_{2},H_{3})$. %
\end{minipage}
}\caption{Setup Algorithm}\label{fig.setup}
\end{figure*}

\begin{figure*}[]
\centering
\fbox{
\begin{minipage}{17.5cm}

\begin{center}{ \sf 
Ticket-Issuer-Reg$\left(\mathcal{I}(x_{i},Y_{I},\tilde{Y}_{I},ID_{I},PP) 
\leftrightarrow\mathcal{CA}(MSK,PP)\right)$}%
\end{center}

\begin{tabular}{lcl}
Ticket Issuer: $\mathcal{I}$ & & Central Authority: $\mathcal{CA}$\\ Select
$x_{i}\stackrel{R}{\leftarrow}\mathbb{Z}_{p}$ and compute
$Y_{I}=\tilde{g}^{x_{i}}$, $\tilde{Y}_{I}=\mathfrak{g}^{x_{i}}$\\ The
secret-public key pair is $(x_{i},Y_{I},\tilde{Y}_{I})$.
&$~~\xrightarrow{ID_{I},Y_{I},\tilde{Y}_{I}}$~~ &  Select
$d_{i},e_{i}\stackrel{R}{\leftarrow}\mathbb{Z}_{p}$ and compute \\ Verify:
$e(\sigma_{I},{Y}_{A}\mathfrak{g}^{e_{i}})\stackrel{?}{=}
e(g_{1}g_{2}^{d_{i}}{Y}_{I},\mathfrak{g})$&
~~$\xleftarrow{\sigma_{I},d_{i},e_{i}}$~~ &
$\sigma_{I}=(g_{1}g_{2}^{d_{i}}{Y}_{I})^{\frac{1}{\alpha+e_{i}}}$.\\ Keep the
credential as $Cred_{I}=(d_{i},e_{i},\sigma_{I})$ & &Store
$(ID_{I},Y_{I},\tilde{Y}_{I},(d_{i},e_{i},\sigma_{I}))$.\\
\end{tabular}
\medskip
\begin{center}{ \sf Ticket-Verifier-Reg$\left(\mathcal{V}(ID_{V},PP)\leftrightarrow\mathcal{CA}(MSK,PP)\right)$} \end{center}
\begin{tabular}{lcl}
Ticket-Verifier: $\mathcal{V}$ & & Central Authority: $\mathcal{CA}$\\
 &$~~\xrightarrow{ID_{V}}$~~ &   Select 
 $d_{v},e_{v}\stackrel{R}{\leftarrow}\mathbb{Z}_{p}$ and compute\\

Verify: 
$e(\sigma_{V},{Y}_{A}\mathfrak{g}^{e_{v}})\stackrel{?}{=}e(g_{1}g_{2}^{d_{v}}\tilde{g}^{H_{1}(ID_{V})},\mathfrak{g})$;
& $\xleftarrow[SK_{V}]{\sigma_{V},d_{v},e_{v}}$ 
& $\sigma_{V}=(g_{1}g_{2}^{d_{v}}\tilde{g}^{H_{1}(ID_{V})})^{\frac{1}{\alpha+e_{v}}}$,\\
$e(\tilde{g},SK_{V})\stackrel{?}{=}e(\tilde{Y}_{A},H_{2}(ID_{V}))$;
& & $SK_{V}=H_{2}(ID_{V})^{\beta}$.\\
Keep the credential as $Cred_{V}=(r_{v},e_{v},\sigma_{V})$ and 
& &
Store $(ID_{V},(d_{v},e_{v},\sigma_{V}),SK_{V})$.\\
 the secret key as $SK_{V}$. 
\end{tabular}
\medskip
\begin{center}{\sf User-Reg$\left(\mathcal{U}(x_{u},Y_{U},ID_{U},PP)\leftrightarrow\mathcal{CA}(MSK,PP)\right)$} \end{center}
\begin{tabular}{lcl}
User: $\mathcal{U}$ & & Central Authority: $\mathcal{CA}$\\
Select $x_{u}\stackrel{R}{\leftarrow}\mathbb{Z}_{p}$, and compute 
$Y_{U}=\tilde{g}^{x_{u}}$ &&\\
This secret-public key pair is $(x_{u},Y_{U})$ & $~~\xrightarrow{ID_{U},Y_{U}}$ 
& Select $d_{u},e_{u}\stackrel{R}{\leftarrow}\mathbb{Z}_{p}$ and compute\\

Verify: 
$e(\sigma_{U},Y_{U}\mathfrak{g}^{e_{u}})\stackrel{?}{=}e(g_{1}g_{2}^{d_{u}}{Y}_{U},\mathfrak{g})$&
 ~~$\xleftarrow{\sigma_{U},d_{u},e_{u}}$~~ & 
$\sigma_{U}=(g_{1}g_{2}^{d_{u}}{Y}_{U})^{\frac{1}{\alpha+e_{u}}}$\\
Keep the credential as $Cred_{U}=(d_{u},e_{u},\sigma_{U})$.& & Store 
$(ID_{U},Y_{U},(d_{u},e_{u},\sigma_{U}))$\\
\end{tabular}
\medskip
\begin{center}{ \sf Central-Verifier-Reg$\left(\mathcal{CV}(x_{cv},Y_{CV},ID_{CV},PP)\leftrightarrow\mathcal{CA}(MSK,PP)\right)$}\end{center}
\begin{tabular}{lcl}
Central Verifier: $\mathcal{CV}$ & & Central Authority: $\mathcal{CA}$\\
Select $x_{cv}\stackrel{R}{\leftarrow}\mathbb{Z}_{p}$, and compute 
$Y_{CV}=\tilde{g}^{x_{cv}}$. &&\\
The secret-public key pair is $(x_{cv},Y_{CV})$ & 
$~~\xrightarrow{ID_{CV},Y_{CV}}$~~  & Select 
$d_{cv},e_{cv}\stackrel{R}{\leftarrow}\mathbb{Z}_{p}$ and compute\\
Verify: 
$e(\sigma_{cv},Y_{A}\mathfrak{g}^{e_{cv}})\stackrel{?}{=}e(g_{1}g_{2}^{d_{cv}}{Y}_{CV},\mathfrak{g})$
 & ~~$\xleftarrow{\sigma_{CV},d_{cv},e_{cv}}$~~ & 
$\sigma_{CV}=(g_{1}g_{2}^{d_{cv}}{Y}_{CV})^{\frac{1}{\alpha+e_{cv}}}$\\
Keep the credential as $Cred_{CV}=(d_{cv},e_{cv},\sigma_{CV})$ & & Store 
$(ID_{CV},Y_{CV},(d_{cv},e_{cv},\sigma_{CV}))$\\
\end{tabular}
\end{minipage}
}\caption{Registration Algorithm}\label{fig.reg}
\end{figure*}

\begin{figure*}[]
\centering
\fbox{
\begin{minipage}{17.5cm}
\begin{center}$\mbox{{\sf Ticket-Issuing}}\left(\mathcal{U}(x_{u},Cred_{U},PP)\leftrightarrow \mathcal{I}(x_{i},Crd_{I},PP)\right)$\end{center}

Suppose that $J_{U}$ is $\mathcal{U}$'s service set consisting of the identities $ID_{V}$ of ticket verifiers and the central verifier $ID_{CV}$.
 \medskip
 
\begin{tabular}{lcl}
{\hspace{1.2cm} User: $\mathcal{U}$} & & \hspace{1.2cm} Ticket Issuer: $\mathcal{I}$\\
\medskip

Compute $A_{U}=g_{1}g_{2}^{d_{u}}{Y}_{U}$. & &\\
Select $y_{1},y_{2},y_{3}\stackrel{R}{\leftarrow}\mathbb{Z}_{p}$ and \\
compute $y_{4}=\frac{1}{y_{1}}$,  $\bar{\sigma}_{U}=\sigma_{U}^{y_{1}}$, & & 
Verify $\prod_{U}^{1}$ and 
$e(\bar{\sigma}_{U},Y_{A})\stackrel{?}{=}e(\tilde{\sigma}_{U},\mathfrak{g})$.\\
$y=d_{u}-y_{2}y_{4}$, $\bar{A}_{U}=A_{U}^{y_{1}}g_{2}^{-y_{2}}$, & & Select 
$r_{u}\stackrel{R}{\leftarrow}\mathbb{Z}_{p}$, and compute 
$R_{U}=\bar{g}^{r_{u}}$.\\
$\tilde{\sigma}_{U}=\bar{\sigma}_{U}^{-e_{u}}A_{U}^{y_{1}}(=\bar{\sigma}_{U}^{\alpha})$,
 $(k_{v}=$ & & For $ID_{V}\in J_{U}$, select 
$t_{v},w_{v},z_{v}\stackrel{R}{\leftarrow}\mathbb{Z}_{p}$, and   compute \\
  $H_{1}(y_{3}||ID_{V}),$ $P_{V}=Y_{U}{Y}_{CV}^{k_{v}},$ & & $D_{V}=H_{3}(R_{U}||{ID_{V}})$,  $E_{V}^{1}=e(\tilde{Y}_{A}, H_{2}(ID_{V}))^{t_{v}}$,\\
   $ Q_{V}=\tilde{g}^{k_{v}})_{ID_{V}\in J_{U}}$.& & $E_{V}^{2}=\tilde{g}^{t_{v}}$,  $E_{V}^{3}=(\vartheta_{1}\vartheta_{2}^{H_{1}(TP||Text_{1}\footnote{$Text_{1}$ specifies the travel time and  other information required by the proxy verification.})})^{t_{v}}$,
\vspace{0.1cm} \\

Compute the proof $\prod_{U}^{1}:$ & 
$\xrightarrow[((P_{V},Q_{V})_{ID_{V}\in J_{U}})]{\bar{\sigma}_{U},\tilde{\sigma}_{U},\bar{A}_{U},J_{U},\prod_{U}^{1}}$ &  $K_{V}=\tilde{g}^{H_{1}(ID_{V})}{Y}_{CV}^{t_{v}}$,
 $s_{v}=H_{1}(P_{V}||Q_{V}||E_{V}^{1}||E_{V}^{2}||E_{V}^{3}$

\\
$\mbox{PoK}\large\{(x_{u},d_{u},e_{u},\sigma_{U},y,y_{1},y_{2},y_{4},$ 
& &  $||K_{V}||Text_{2} \footnote{$Text_{2}$ consists of the system version information and all other information which can be used by verifiers to validate the ticket, e.g. valid period, ticket type, {\em etc}.})$ and  $Z_{V}=(g_{1}g_{2}^{w_{v}}g_{3}^{s_{v}})^{\frac{1}{x_{i}+z_{v}}}$.
\\
$(k_{v})_{ID_{V}\in J_{U}}): \frac{\tilde{\sigma}_{U}}{\bar{A}_{U}}=\bar{\sigma}_{U}^{-e_{u}}g_{2}^{y_{2}}$ 
& & The authentication tag is  $Tag_{V}=((P_{V},Q_{V}),(E_{V}^{1},$
\\
$ \wedge~ g_{1}^{-1}= \bar{A}_{U}^{-y_{4}}g_{2}^{y}\tilde{g}^{x_{u}}$ $\wedge~$ $(P_{V}= $
 & &      $E_{V}^{2},E_{V}^{3},K_{V},Text_{1},Text_{2}),(s_{v},w_{v},z_{v},Z_{V}))$, 
 \\
 $\tilde{g}^{x_{u}}{Y}_{CV}^{k_{v}} \wedge Q_{V}=\tilde{g}^{k_{v}})_{ID_{V}\in J_{U}}\large\}$ 
 & &   where    $s_{v}$ is  the serial numbers of $Tag_{V}$.     \\

& & Select $w,z\stackrel{R}{\leftarrow}\mathbb{Z}_{p}$ and compute\\
& & $s=H_{1}(s_{1}||s_{2}||\cdots||s_{|J_{U}|})$ and $Z=(g_{1}g_{2}^{w}g_{3}^{s})^{\frac{1}{x_{i}+z}}$\\
& & where $s$ is the serial number of the ticket.\\
For $ID_{V}\in J_{U}$, verify 
& $\xleftarrow{R_{U},T_{U}}$ 
& The ticket is: $T_{U}=\big\{(D_{V},Tag_{V})|ID_{V}\in J_{U}\big\} \cup \big\{(s,$\\
$D_{V}\stackrel{?}{=}H_{3}(R_{U}||ID_{V})$,
& &  
$w,z,Z)\big\}$. \\
$s_{v}\stackrel{?}{=}H_{1}(P_{V}||Q_{V}||E_{V}^{1}||E_{V}^{2}||E_{V}^{3}||K_{V}$
& & \\ 
~~~~~~~~~~~$||Text_{2})$. & & \\
$s\stackrel{?}{=}H_{1}(s_{1}||s_{2}||\cdots||s_{|J_{U}|})$, 
& & \\
$e(Z_{V},\tilde{Y}_{I}\mathfrak{g}^{z_{v}})\stackrel{?}{=}e(g_{1}g_{2}^{w_{v}}g_{3}^{s_{v}},\mathfrak{g})$. & &\\
and $e(Z,\tilde{Y}_{I}\mathfrak{g}^{z})=e(g_{1}g_{2}^{w}g_{3}^{s},\mathfrak{g})$& &\\
Keep  $(x_{3},R_{U})$ secret.\\
\end{tabular}
\end{minipage}
}\caption{Ticket Issuing Algorithm}\label{fig.t-i}
\end{figure*}

\begin{figure*}[]
\centering
\fbox{
\begin{minipage}{17.5cm}
\begin{center}$\mbox{{\sf Ticket-Validation}}\left(\mathcal{U}(x_{u},Tag_{V},PP)\leftrightarrow \mathcal{V}(ID_{V},PP)\right)$\end{center}
\begin{tabular}{lcl}
User: $\mathcal{U}$ && Ticket verifier: $\mathcal{V}$ $(ID_{V}\in J_{U})$\\
Compute $D_{V}=H_{3}(R_{U}||ID_{V})$ & $\xleftarrow{{ID_{V}}}$ & 
Initialize 
a table $T_{V}$.\\
and search $(D_{V},Tag_{V}).$\\
Compute $k_{v}=H_{1}(y_{3}||ID_{V})$  \\
and the proof: $\prod_{U}^{2}:$ & & If  $(s_{v},w_{v},z_{v},Z_{V})\in T_{V}$, 
abort; otherwise, add $(s_{v},w_{v},z_{v},Z_{V})$ in\\
$\mbox{PoK}\{(x_{u},z_{v}): P_{V}=\tilde{g}^{x_{u}}{Y}_{CV}^{k_{v}}~\wedge$ &~~ 
$\xrightarrow{\prod_{U}^{2},Tag_{V}}$~~& $T_{V}$ and go to the next step. \\
$~~~~~~~~~~~~~~~~~~~~~~~Q_{V}=\tilde{g}^{k_{v}}\}$. & & Check:\\

& & (1) The correctness of $\prod_{U}^{2}$;\\
& & (2) $s_{v}\stackrel{?}{=}H_{1}(P_{V}||Q_{V}||E_{V}^{1}||E_{V}^{2}||E_{V}^{3}||K_{V}||Text_{2})$;\\
& & (3) $e(E_{V}^{2},SK_{V})\stackrel{?}{=}E_{V}^{1}$;\\
& & (4)  $e(Z_{V},Y_{I}\mathfrak{g}^{z_{v}})\stackrel{?}{=}e(g_{1}g_{2}^{w_{v}}g_{3}^{s_{v}},\mathfrak{g})$.\\
& & If (1), (2), (3) and (4)  hold, the ticket is valid; otherwise, it is invalid.\\

\end{tabular}
\end{minipage}
}\caption{Ticket Validation Algorithm}\label{fig.t-v}
\end{figure*}

\begin{figure*}[]
\centering
\fbox{
\begin{minipage}{17.5cm}
\begin{center}{\sf Ticket-Trace}$(x_{cv},T_{U},PP)$\end{center} 

Given a ticket $T_{U}$, $\mathcal{CV}$ works as follows:
\medskip

 (1) Let $\Omega_{U}=\{\}$. For each $Tag_V$ in $T_U$: 
 
 \hspace{0.3cm} a) Compute: $Y_{U}=\frac{P_{V}}{Q_{V}^{x_{cv}}}$ and $g^{H_{1}(ID_{V})}=\frac{K_{V}}{(E_{V}^{2})^{x_{cv}}}$; b) Look up $g^{H_{1}(ID_V)}$ and $\mathcal{V}$'s identity.
\medskip

\hspace{0.3cm} Check: 
\medskip

\hspace{0.3cm} (c1) $s_{v}\stackrel{?}{=}H_{1}(P_{V}||Q_{V}||E_{V}^{1}||E_{V}^{2}||E_{V}^{3}||K_{V}||Text_{1})$;  (c2) $e(Z_{V},Y_{I}\mathfrak{g}^{z_{v}})\stackrel{?}{=}e(g_{1}g_{2}^{w_{v}}g_{3}^{s_{v}},\mathfrak{g})$;

 \hspace{0.3cm} (d) If (c1) and (c2) hold, set $\Omega_U=\Omega_U\cup\{ID_{V}\}$; otherwise abort. 
 
\hspace{0.3cm}  (e) Verify $Y_U$ remains the same for all tags.\\
 \medskip
  (2) $s\stackrel{?}{=}H_{1}(s_{1}||s_{2}||\cdots||s_{|J_{U}|})$;\\
  \medskip
 (3) $e(Z,\tilde{Y}_{I}\mathfrak{g}^{z})\stackrel{?}{=}e(g_{1}g_{2}^{w}g_{3}^{s},\mathfrak{g})$.\\
   If (1), (2) and (3) hold, $\mathcal{CV}$ can determine that the service    
  information of $\mathcal{U}$ with public key $Y_{U}$ is:  $J_{U}=\Omega_U$;
  otherwise, the trace has failed.

\end{minipage}
}\caption{Ticket Trace Algorithm}\label{fig.t-t}
\end{figure*}

\begin{figure*}[]
\centering
\fbox{
\begin{minipage}{17.5cm}
\begin{center}$\mbox{{\sf Proxy-Key-Generation}}\left(\mathcal{V}(ID_{V})\leftrightarrow \mathcal{CA}(\alpha,ID_{V},ID_{V'})\right)$\end{center}
\begin{tabular}{lcl}
Verifier: $\mathcal{V'}$ && Central Verifier: $\mathcal{CA}$ \\
& & If there is a disruption on the verifier $\mathcal{V}$  and users  should go through verifier $\mathcal{V}'$.\\
& & $\mathcal{CA}$ works as follows:\\
& &(1) check the registration information and find $ 
(ID_{V},(d_{v},e_{v},\sigma_{V}),SK_{V})$ and\\
& &~~~ $(ID_{V'},(d_{v'},e_{v'},\sigma_{V'}),SK_{V'})$;\\
& & (2) choose $\beta_{v}\stackrel{R}{\leftarrow}\mathbb{Z}_{p}$ and compute 
$RK_{1}=\tilde{g}^{\beta_{v}}, 
RK_{2}=(\vartheta_{1}\vartheta_{2}^{H_{1}(TP||Text_{1})})^{\beta_{v}}\cdot 
\frac{SK_{V}}{SK_{V'}}$\\
Keep  $RK_{\mathcal{A}\rightarrow\mathcal{A}'}$.
& $\xleftarrow{RK_{\mathcal{A}\rightarrow\mathcal{A}'}}$ &
~~~which is bound  with the  time period $TP$ and system requirements. The re-key is \\
& &~~ $RK_{\mathcal{A}\rightarrow\mathcal{A}'}=(RK_{1},RK_{2})$. \\
& & $\mathcal{CA}$ publishes the proxy information $(ID_{V},ID_{V'})$.\\
\end{tabular}
\medskip

\end{minipage}
}\caption{Proxy Key Generation Algorithm}\label{fig.p-k-g}
\end{figure*}

\begin{figure*}[]
\centering
\fbox{
\begin{minipage}{17.5cm}
\begin{center}$\mbox{{\sf Proxy-Ticket-Validation}}\left(\mathcal{U}(x_{u},Tag_{V},PP)\leftrightarrow \mathcal{V'}(ID_{V'},SK_{V'},RK_{\mathcal{A}\rightarrow\mathcal{A}'},PP)\right)$\end{center}
\begin{tabular}{lcl}
User: $\mathcal{U}$ && Ticket verifier: $\mathcal{V'}$ $(ID_{V'}\notin J_{U})$\\
Compute $D_{V}=H_{2}(C_{U}||ID_{V})$ & $\xleftarrow{ID_{V'}}$ & Initialize a 
table $T_{V'}$.\\
and search $(D_{V},Tag_{V}).$\\
Compute $k_{v}=H_{1}(z_{u}||ID_{V})$  \\
and the proof: $\prod_{U}^{2}:$\\
$\mbox{PoK}\{(x_{u},z_{v}): P_{V}=\tilde{g}^{x_{u}}{Y}_{CV}^{k_{v}}~\wedge$ &~~ 
$\xrightarrow{\prod_{U}^{2},Tag_{V}}$~~&  If  $(s_{v},w_{v},z_{v},Z_{V})\in 
T_{V'}$, abort; otherwise, add $(s_{v},w_{v},z_{v},Z_{V})$ in \\
$~~~~~~~~~~~~~~~~~~~~~~~Q_{V}=\tilde{g}^{z_{v}}\}$. & & $T_{V'}$ and go to the 
next step. \\
& & Compute: \\
& &  $\Theta_{1}=RK_{2}\cdot SK_{V'}=(\vartheta_{1}\vartheta_{2}^{H_{1}(TP||Text_{1})})^{\beta_{v}}\cdot H_{2}(ID_{V})^{\alpha}$ and \\
&  & $\Theta_{2}=\frac{e(E_{V}^{2},\Theta_{1})}{e(RK_{1},E_{V}^{3},)}$.\\
& & Check:\\
& & (1) The correctness of $\prod_{U}^{2}$; (2)  $s_{v}\stackrel{?}{=}H_{1}(P_{V}||Q_{V}||E_{V}^{1}||E_{V}^{2}||E_{V}^{3}||K_{V}||$\\
& & $Text_{2})$; (3) $\Theta_{2}\stackrel{?}{=}E_{V}^{1}$; (4)  $e(Z_{V},Y_{I}\mathfrak{g}^{z_{v}})\stackrel{?}{=}e(g_{1}g_{2}^{w_{v}}g_{3}^{s_{v}},\mathfrak{g})$.\\
& & If (1), (2), (3) and (4)  hold, the ticket is valid;  otherwise, it is invalid.\\
\end{tabular}
\medskip

{\em Note: To prevent double spend a tag/ticket: (1) a central server/database is required to store the verification records and can be accessed by any verifier in the systems; or 
(2) $\mathcal{V}$ should send the verification records on the disruption day to $\mathcal{V}'$ and $\mathcal{V}'$ should send back the proxy verification records on the disruption day to $\mathcal{V}$.}
\end{minipage}
}\caption{Proxy Ticket Validation Algorithm}\label{fig.p-t-v}
\end{figure*}

\subsection{Correctness}
The details of the zero-knowledge proofs of 
$\prod_{U}^{1}$ and $\prod_{U}^{2}$ are provided in in Appendix \ref{d1} and Appendix \ref{d2}.
\medskip

Our scheme is correct as the following equations hold.
In the registration algorithm, the credentials can be verified by the following equations.
\begin{equation*}
\begin{split}
e(\sigma_{I},Y_{A}\mathfrak{g}^{e_{i}})=e((gh^{d_{i}}Y_{I})^{\frac{1}{\alpha+e_{i}}},\mathfrak{g}^{\alpha+e_{i}})=e(gh^{d_{i}}Y_{I},\mathfrak{g}),
\end{split}
\end{equation*}
\begin{equation*}
\begin{split}
e(\sigma_{V},Y_{A}\mathfrak{g}^{e_{v}})=e((gh^{d_{v}}Y_{V})^{\frac{1}{\alpha+e_{v}}},\mathfrak{g}^{\alpha+e_{v}})=e(gh^{d_{v}}Y_{V},\mathfrak{g}),
\end{split}
\end{equation*}
\begin{equation*}
\begin{split}
e(\sigma_{U},Y_{A}\mathfrak{g}^{e_{u}})=e((gh^{d_{u}}Y_{U})^{\frac{1}{\alpha+e_{u}}},\mathfrak{g}^{\alpha+e_{u}})=e(gh^{d_{u}}Y_{U},\mathfrak{g})
\end{split}
\end{equation*}
and
\begin{equation*}
\begin{split}
 e(\sigma_{CV},Y_{A}\mathfrak{g}^{e_{cv}}) & =e((gh^{d_{cv}}Y_{CV})^{\frac{1}{\alpha+e_{cv}}},\mathfrak{g}^{\alpha+e_{cv}})\\
& =e(gh^{d_{cv}}Y_{CV},\mathfrak{g}),
\end{split}
\end{equation*}

In the ticket issuing algorithm, the correctness of the zero knowledge proof $\prod_{U}^{1}$ and the ticket can be verified by the following equations.
\begin{equation*}
\begin{split}
\tilde{\sigma}_{U}&=\bar{\sigma_{U}}^{-e_{u}}A_{U}^{y_{1}}=\sigma_{U}^{-e_{u}y_{1}}A_{U}^{y_{1}}=A_{U}^{\frac{-e_{u}y_{1}}{\alpha+e_{u}}}A_{U}^{y_{1}}\\
&=A_{U}^{\frac{-y_{1}(e_{u}+\alpha)+y_{1}\alpha}{\alpha+e_{u}}}Y_{U}^{y_{1}}=A_{U}^{-y_{1}}A_{U}^{\frac{y_{1}\alpha}{\alpha+e_{u}}}A_{U}^{y_{1}}\\
&=(A_{U}^{\frac{1}{\alpha+e_{u}}})^{y_{1}\alpha}=(\sigma_{U}^{y_{1}})^{\alpha}=\bar{\sigma}_{U}^{\alpha},
\end{split}
\end{equation*}

\begin{equation*}
\begin{split}
\frac{\tilde{\sigma}_{U}}{\bar{A}_{U}}=\frac{\bar{\sigma}_{U}^{-e_{u}}A_{U}^{y_{1}}}{A_{U}^{y_{1}}g_{2}^{-v_{2}}}=\bar{\sigma}_{U}^{-e_{u}}g_{2}^{v_{2}},
\end{split}
\end{equation*}

\begin{equation*}
\begin{split}
\bar{A}_{U}^{-y_{3}}\tilde{g}^{x_{u}}g_{2}^{y}& =(A_{U}^{y_{1}}g_{2}^{-y_{2}})^{-y_{4}}\tilde{g}^{x_{u}}g_{2}^{y}\\
&=((g_{1}g_{2}^{d_{u}}Y_{U})^{y_{1}}g_{2}^{-y_{2}})^{-y_{4}}\tilde{g}^{x_{u}}g_{2}^{y}\\
&=(g_{1}g_{2}^{d_{u}}Y_{U})^{-1}g_{2}^{y_{2}y_{4}}\tilde{g}^{x_{u}}g_{2}^{y}\\
&=g^{-1}g_{2}^{-d_{u}}Y_{U}^{-1}Y_{U}g_{2}^{y+y_{2}y_{4}}\\
&=g^{-1}g_{2}^{y+y_{2}y_{3}-d_{u}}\\
&=g^{-1},
\end{split}
\end{equation*}
\begin{equation*}
\begin{split}
e(Z_{V},\tilde{Y}_{I}\mathfrak{g}^{z_{v}})&=e((g_{1}g_{2}^{w_{v}}g_{3}^{s_{v}})^{\frac{1}{x_{i}+z_{v}}},\mathfrak{g}^{x_{i}+z_{v}})\\
&=e(g_{1}g_{2}^{w_{v}}g_{3}^{s_{v}},\mathfrak{g})
\end{split}
\end{equation*}
and
\begin{equation*}
\begin{split}
e(Z_{CV},\tilde{Y}_{I}\mathfrak{g}^{z_{cv}})&=e((g_{1}g_{2}^{w_{cv}}g_{3}^{s_{cv}})^{\frac{1}{x_{i}+z_{cv}}},\mathfrak{g}^{x_{i}+z_{cv}})\\
&=e(g_{1}g_{2}^{w_{cv}}g_{3}^{s_{cv}},\mathfrak{g}).
\end{split}
\end{equation*}

In the ticket validation algorithm, a tag can be validated by the following equations.
\begin{equation*}
\begin{split}
e(E_{V}^{2},SK_{V})& =e(\tilde{g}^{t_{v}},H_{2}(ID_{V}^{\beta}))=e(\tilde{g}^{\beta},H_{2}(ID_{V})^{t_{v}})\\
& =e(\tilde{Y}_{A},H_{2}(ID_{V}))^{t_{v}}=E_{V}^{1},
\end{split}
\end{equation*}
\begin{equation*}
\begin{split}
\frac{P_{V}}{Q_{V}^{x_{cv}}}=\frac{Y_{U}Y_{CV}^{k_{v}}}{\tilde{g}^{x_{cv}k_{v}}}=\frac{Y_{U}Y_{CV}^{k_{v}}}{Y_{CV}^{k_{v}}}=Y_{U}
\end{split}
\end{equation*}
and
\begin{equation*}
\begin{split}
\frac{T_{V}}{(E_{V}^{2})^{x_{cv}}}=\frac{\tilde{g}^{H_{1}(ID_{V})}Y_{CV}^{t_{v}}}{\tilde{g}^{x_{cv}t_{v}}}=\frac{\tilde{g}^{H_{1}(ID_{V})}Y_{CV}^{t_{v}}}{Y_{CV}^{t_{v}}}=\tilde{g}^{H_{1}(ID_{V})}.
\end{split}
\end{equation*}
In the proxy ticket validation algorithm, a tag can be validated by a proxy since the following equations hold.
\begin{equation*}
\begin{split}
\Theta_{1}& =RK_{2}\cdot SK_{V'}=(\vartheta_{1}\vartheta_{2}^{H_{1}(TP||Text_{1})})^{\beta_{v}}\cdot \frac{SK_{V}}{SK_{V'}}\cdot SK_{V'}\\
& =(\vartheta_{1}\vartheta_{2}^{H_{1}(TP||Text_{1})})^{\beta_{v}}\cdot  SK_{V}
\end{split}
\end{equation*}
and
\begin{equation*}
\begin{split}
\Theta_{2}& =\frac{e(E_{V}^{2},\Theta_{1})}{e(RK_{1},E_{V}^{3})}=\frac{e(\tilde{g}^{t_{v}},(\vartheta_{1}\vartheta_{2}^{H_{1}(TP||Text_{1})})^{\beta_{v}}\cdot  SK_{V})}{e(\tilde{g}^{\beta_{v}},(\vartheta_{1}\vartheta_{2}^{H_{1}(TT||Text_{1})})^{t_{v}})}\\
&=\frac{e(\tilde{g}^{\beta_{v}},(\vartheta_{1}\vartheta_{2}^{H_{1}(TP||Text_{1})})^{t_{v}})\cdot e(\tilde{g}^{t_{v}},H_{2}(ID_{V})^{\beta})}{e(\tilde{g}^{\beta_{v}},(\vartheta_{1}\vartheta_{2}^{H_{1}(TP||Text_{1})})^{t_{v}})}\\
&=e(\tilde{g}^{\beta},H_{2}(ID_{V}))^{t_{v}}=e(\tilde{Y}_{A},H_{2}(ID_{V}))^{t_{v}}=E_{V}^{1}.
\end{split}
\end{equation*}

\section{Security Analysis}\label{security_analysis}
In this section, the security of our scheme 
is formally 
proven. 

\begin{theorem}\label{theorem:unforg}%
Our scheme described in Fig. \ref{fig.setup}, Fig. \ref{fig.reg}, Fig.
\ref{fig.t-i}, Fig. \ref{fig.t-v}, Fig. \ref{fig.t-t}, Fig. \ref{fig.p-k-g} and
Fig. \ref{fig.p-t-v} is $(\varrho, \epsilon'(\ell))$-unforgeable if and
only if the $(q,\epsilon(\ell))$-JoC-q-SDH assumption holds on the bilinear group
$(e,p,\mathbb{G}_{1},\mathbb{G}_{2},\mathbb{G}_{\tau})$ and
$H_{1},H_{2}$ and $H_{3}$ are secure cryptographic hash functions, where
$\varrho$ is the number of ticket issuing queries made by the adversary
$\mathcal{A}$, $\varrho<q$ and $\epsilon(\ell)
\geq(\frac{p-q}{p}+\frac{1}{p}+\frac{p-1}{p^{3}})\epsilon'(\ell)$.
\end{theorem}

\begin{IEEEproof}
Suppose that there exists an adversary $\mathcal{A}$ that can break the unforgeability of our scheme, we can construct an algorithm $\mathcal{B}$ which can use $\mathcal{A}$ to break the {\em JoC-q-SDH} assumption. Given  a $(q+3)$-tuple 
$(g,g^{x},\cdots,g^{x^{q}},\mathfrak{g},
\mathfrak{g}^{x})\in\mathbb{G}_{1}^{q+1}\times\mathbb{G}_{2}^{2}$, $\mathcal{B}$ will output $(c,g^{\frac{1}{x+c}})\in\mathbb{Z}_{p}\times\mathbb{G}_{1}$ where $c\neq x$. 
\medskip

\noindent{\sf Setup.} $\mathcal{B}$ selects $\pi_{1},\pi_{2},\cdots,\pi_{q-1}\stackrel{R}{\leftarrow}\mathbb{Z}_{p}$, and sets $f(x)=\prod_{i=1}^{q-1}(x+\pi_{i})=\sum_{i=0}^{q-1}\theta_{i}x^{i}$, $f_{i}(x)=\frac{f(x)}{x+\pi_{i}}=\sum_{j=0}^{q-2}\omega_{i_{j}}x^{j}$, $\tilde{g}=\prod_{i=0}^{q-1}(g^{x^{i}})^{\chi_{i}}=g^{f(x)}$, $\hat{g}=\prod_{i=0}^{q-1}(g^{x^{i+1}})^{\theta_{i}}=\tilde{g}^{x}$.  $\mathcal{B}$ selects $\pi,a,k\stackrel{R}{\leftarrow}\mathbb{Z}_{p}$ and computs $g_{2}=((\hat{g}\bar{g}^{\pi})^{k}\bar{g}^{-1})^{\frac{1}{a}}=\bar{g}^{\frac{(x+\pi)k-1}{a}}$.  $\mathcal{B}$ 
 selects  $\gamma_{1},\gamma_{2},\gamma_{3}\stackrel{R}{\leftarrow}\mathbb{Z}_{p}$  and  computes $g_{1}=\bar{g}^{\gamma_{1}}$, $g_{3}=g_{2}^{\gamma_{2}}$ and $\bar{g}=g^{\gamma_{3}}$. $\mathcal{B}$ selects hash functions $H_{1}:\{0,1\}^{*}\rightarrow\mathbb{Z}_{p}$, $H_{2}:\{0,1\}\rightarrow \mathbb{G}_{2}$ and $H_{3}:\{0,1\}^{*}\rightarrow \{0,1\}^{\ell'} ~(\ell'\leq\ell)$. $\mathcal{B}$ selects $\alpha,\beta,\stackrel{R}{\leftarrow}\mathbb{Z}_{p}$ and $\vartheta_{1},\vartheta_{2}\stackrel{R}{\leftarrow}\mathbb{G}_{2}$. $\mathcal{B}$ computes ${Y}_{A}=\mathfrak{g}^{\alpha}$ and  $\tilde{Y}_{A}=\tilde{g}^{\beta}$. The master secret key is $MSK=(\alpha,\beta)$ and the public parameters are $PP=(e,p,\mathbb{G}_{1},\mathbb{G}_{2},\mathbb{G}_{\tau},\tilde{g},\bar{g},g_{1},g_{2},g_{3},\mathfrak{g},\vartheta_{1},\vartheta_{2},Y_{A},\tilde{Y}_{A},$ $H_{1},H_{2},H_{3})$.  
\medskip

\noindent{\sf Registration Query.} $\mathcal{A}$ adaptively makes the following queries:
\begin{enumerate}
\item{\em Ticket Issuer Registration Query.} $\mathcal{B}$ sets $Y_{I}=\tilde{g}^{x}$ and $\tilde{Y}_{I}=\mathfrak{g}^{x}$. $\mathcal{B}$ selects $d_{i},e_{i}\stackrel{R}{\leftarrow}\mathbb{Z}_{p}$ and computes $\sigma_{I}=(g_{1}g_{2}^{d_{i}}Y_{I})^{\frac{1}{\alpha+e_{i}}}$. $\mathcal{B}$ sends $(\sigma_{I},Y_{I},\tilde{Y}_{I})$ to $\mathcal{A}$.

\item{\em Ticket Verifier Registration Query.} $\mathcal{A}$ submits an identity $ID_{V}\in\{0,1\}^{*}$. $\mathcal{B}$ selects $d_{v},e_{v}\stackrel{R}{\leftarrow}\mathbb{Z}_{p}$, and computes $\sigma_{V}=(g_{1}g_{2}^{d_{v}}\tilde{g}^{H_{1}(ID_{V})})^{\frac{1}{\alpha+e_{v}}}$ and $SK_{V}=H_{2}(ID_{V})^{\beta}$. $\mathcal{B}$ sends $(\sigma_{V},SK_{V})$ to $\mathcal{A}$.

\item{\em User Registration Query.} $\mathcal{A}$ submits an identity $ID_{U}$ and the corresponding public key $Y_{U}$. $\mathcal{C}$ selects $e_{u},d_{u}\stackrel{R}{\leftarrow}\mathbb{Z}_{p}$ and computes $\sigma_{U}=(g_{1}g_{2}^{d_{u}}Y_{U})^{\frac{1}{\alpha+e_{u}}}$. $\mathcal{B}$ returns $\sigma_{U}$ to $\mathcal{A}$.

\item{\em Central Verifier Registration Query.} $\mathcal{A}$ submits an identity $ID_{CV}\in\{0,1\}^{*}$ and the corresponding public key $Y_{CV}$. $\mathcal{B}$ selects $d_{cv},e_{cv}\stackrel{R}{\leftarrow}\mathbb{Z}_{p}$ and computes $\sigma_{CV}=(g_{1}g_{2}^{d_{v}}Y_{CV})^{\frac{1}{\alpha+e_{cv}}}$. $\mathcal{B}$ returns $\sigma_{CV}$ to $\mathcal{A}$.
\end{enumerate}

\noindent{\sf Ticket Issuing Query.}  $\mathcal{A}$ adaptively submits  a set of service information $J_{U}$ and a set of pseudonyms $(P_{V},Q_{V})$ for $ID_{V}\in J_{U}$ and a proof $\prod_{U}^{1}: \mbox{PoK}\LARGE\{(x_{u},d_{u},e_{u},\sigma_{U},y,y_{1},y_{2},$ $y_{4},(k_{v})_{{V}\in J_{U}}): \frac{\tilde{\sigma}_{U}}{\bar{A}_{U}}=\bar{\sigma}_{U}^{-e_{u}}g_{2}^{y_{2}} 
 \wedge~ g_{1}^{-1}= \bar{A}_{U}^{-y_{4}}g_{2}^{y}\tilde{g}^{x_{u}}\wedge~(P_{V}= \tilde{g}^{x_{u}}{Y}_{CV}^{k_{v}} \wedge Q_{V}=\tilde{g}^{k_{v}})_{{V}\in J_{U}}\LARGE\}$. $\mathcal{B}$ checks the proof $\prod_{U}^{1}$ and  $e(\bar{\sigma}_{U},Y_{U})\stackrel{?}{=}e(\tilde{\sigma}_{U},\mathfrak{g})$. If each of them is incorrect, $\mathcal{B}$ aborts; otherwise, $\mathcal{B}$ goes to the next step. $\mathcal{B}$ chooses $r_{u}\stackrel{R}{\leftarrow}\mathbb{Z}_{p}$, and computes $R_{u}=\bar{g}^{r_{u}}$.  For $ID_{V}\in J_{U}$ and let $f_{v}(x)=\frac{f(x)}{x+\pi_{v}}=\sum_{j=0}^{q-2}\omega_{v_{j}}x^{j}$, $\mathcal{B}$ selects $t_{v},w_{v},z_{v}\stackrel{R}{\leftarrow}\mathbb{Z}_{p}$, and computes $D_{V}=H_{3}(R_{U}||ID_{V})$, $E_{V}^{1}=e(\tilde{Y}_{A},H_{2}(ID_{V}))^{t_{v}}$, $E_{V}^{2}=\tilde{g}^{t_{v}}$, $E_{V}^{3}=(\vartheta_{1}\vartheta_{2}^{H_{1}(TP||Text_{1})})^{t_{v}}$, $K_{V}=\tilde{g}^{h_{1}(ID_{V})}Y_{CV}^{t_{v}}$, $s_{v}=H_{1}(P_{V}||Q_{V}||E_{V}^{1}||E_{V}^{2}||E_{V}^{3}||K_{V}||Test_{1}||Test_{2})$ and $Z_{V}=\prod_{j=0}^{q-2}(g^{x^{j}})^{\omega_{v_{j}}(\gamma_{1}+\frac{(\pi k-1)(w_{v}+\gamma_{2}s_{v})}{a})}\prod_{j=0}^{q-2}(g^{x^{j+1}})^{\frac{\omega_{v_{j}}k (w_{v}+\gamma_{2}s_{v})}{a}}$.
\medskip

We claim that $(w_{v},\pi_{v},Z_{V})$ is a valid BBS+ signature on $s_{v}$.

\begin{equation*}
\begin{split}
Z_{V}&=\prod_{j=0}^{q-2}(g^{x^{j}})^{\omega_{v_{j}}(\gamma_{1}+\frac{(\pi k-1)(w_{v}+\gamma_{2}s_{v})}{a})}\prod_{j=0}^{q-2}(g^{x^{j+1}})^{\frac{\omega_{v_{j}}k (w_{v}+\gamma_{2}s_{v})}{a}}\\
& =\prod_{j=0}^{q-2}(g^{\omega_{v_{j}}x^{j}})^{(\gamma_{1}+\frac{(\pi k-1)(w_{v}+\gamma_{2}s_{v})}{a})}\prod_{j=0}^{q-2}(g^{\omega_{v_{j}}x^{j}})^{\frac{xk (w_{v}+\gamma_{2}s_{v})}{a}}\\
&=(g^{f_{v}(x)})^{(\gamma_{1}+\frac{(\pi k-1)(w_{v}+\gamma_{2}s_{v})}{a})}(g^{f_{v}(x)})^{\frac{xk (w_{v}+\gamma_{2}s_{v})}{a}}\\
&=(g^{f(x)})^{(\gamma_{1}+\frac{(\pi k-1)(w_{v}+\gamma_{2}s_{v})}{a})\frac{1}{(x+\pi_{v})}}(g^{f(x)})^{\frac{xk (w_{v}+\gamma_{2}s_{v})}{a(x+\pi_{v})}}\\
&= (g^{f(x)})^{(\gamma_{1}+\frac{(\pi k-1)(w_{v}+\gamma_{2}s_{v})}{a})\frac{1}{(x+\pi_{v})}}(g^{f(x)})^{\frac{xk (w_{v}+\gamma_{2}s_{v})}{a(x+\pi_{v})}}\\
&=((g^{f(x)})^{\gamma_{1}}(g^{f(x)})^{\frac{(\pi k-1)(w_{v}+\gamma_{2}s_{v})+xk (w_{v}+\gamma_{2}s_{v})}{a}})^{\frac{1}{x+\pi_{v}}}\\
&=(\bar{g}^{\gamma_{1}}(\bar{g}^{\frac{(\pi k-1+xk)}{a}})^{w_{v}}\bar{g}^{\frac{(\pi k-1+x k)\gamma_{2}s_{v}}{a}})^{\frac{1}{x+\pi_{v}}}\\
&=(\bar{g}^{\gamma_{1}}(\bar{g}^{\frac{(\pi+x) k-1}{a}})^{w_{v}}\bar{g}^{\frac{((\pi+x) k-1)\gamma_{2}s_{v}}{a}})^{\frac{1}{x+\pi_{v}}}\\
&=(g_{1}g_{2}^{w_{v}}g_{2}^{\gamma_{2}s_{v}})^{\frac{1}{x+\pi_{v}}}=(g_{1}g_{2}^{w_{v}}g_{3}^{s_{v}})^{\frac{1}{x+\pi_{v}}}.
\end{split}
\end{equation*}

Let $f_{cv}=\frac{f(x)}{x+\pi_{cv}}=\sum_{j=0}^{q-2}\omega_{c_{j}}x^{j}$ with $\pi_{cv}\in\{\pi_{1},\pi_{2},\cdots,\pi_{q-1}\}$. $\mathcal{B}$ selects $w_{cv}\stackrel{R}{\leftarrow}\mathbb{Z}_{p}$ and computes $s_{cv}=H_{1}(s_{1}||s_{2}||\cdots||s_{|J_{U}|})$ and $Z_{CV}=\prod_{j=0}^{q-2}(g^{x^{j}})^{\omega_{c_{j}}(\gamma_{1}+\frac{(\pi k-1)(w_{cv}+\gamma_{2}s_{cv})}{a})}\prod_{j=0}^{q-2}(g^{x^{j+1}})^{\frac{\omega_{c_{j}}k (w_{cv}+\gamma_{2}s_{cv})}{a}}$.

When the $q$-th signature is required, $\mathcal{B}$ sets $w_{cv}=a\gamma_{1}-s_{cv}\gamma_{2}$ and $Z_{CV}=\bar{g}^{\gamma_{1}k}$. We claim that $(w_{cv},\pi,Z_{CV})$ is a valid BBS+ signature on $s_{cv}$ since we have
\begin{equation*}
\begin{split}
Z_{CV}&=\bar{g}^{\gamma_{1}k}=(\bar{g}^{\gamma_{1}}\bar{g}^{\frac{a\gamma_{1}(k(x+\pi)-1)}{a}})^{\frac{1}{x+\pi}}\\
&=(\bar{g}^{\gamma_{1}}\bar{g}^{\frac{(w_{cv}+s_{cv}\gamma_{2})(k(x+\pi)-1)}{a}})^{\frac{1}{x+\pi}}\\
&=(\bar{g}^{\gamma_{1}}\bar{g}^{\frac{w_{cv}(k(x+\pi)-1)}{a}}\bar{g}^{\frac{s_{cv}\gamma_{2}(k(x+\pi)-1)}{a}})^{\frac{1}{x+\pi}}\\
&=(\bar{g}^{\gamma_{1}}(\bar{g}^{\frac{(k(x+\pi)-1)}{a}})^{w_{cv}}((\bar{g}^{\frac{(k(x+\pi)-1)}{a}})^{\gamma_{2}})^{s_{cv}})^{\frac{1}{x+\pi}}\\
&=(g_{1}g_{2}^{w_{cv}}g_{3}^{s_{cv}})^{\frac{1}{x+\pi}}
\end{split}
\end{equation*}
The ticket is $T_{U}=\{(D_{V},Tag_{V})|ID_{V}\in J_{U}\}\cup\{s_{cv},w_{cw},$ $\pi_{cv},Z_{CV}\}$. Let $TQ$ be a set consisting of the ticket information queried by $\mathcal{A}$ and initially empty. $\mathcal{B}$ adds $(ID_{U},T_{U},Text_{1},Text_{2})$ into $TQ$ and sends it to $\mathcal{A}$.
\medskip

\noindent{\sf Output.} $\mathcal{A}$ outputs a ticket $T_{U}^{*}=\{(D_{V}*,Tag_{V^{*}})|ID_{V^{*}}\in J_{U}\}\cup\{s_{cv}^{*},w_{cv}^{*},z_{cv}^{*},Z_{CV}^{*}\}$. Suppose that $(s^{*},w^{*},z^{*},Z^{*})\in \{Tag_{V^{*}}|ID_{V^{*}\in J_{U^{*}}}\}\cup \{s_{cv}^{*},w_{cv}^{*},z_{cv}^{*},Z_{CV}^{*}\}$ is a forged tag.

The following three cases are considered.

\begin{itemize}%
\item{\em Case-1.} $z^{*}\notin\{\pi_{1},\pi_{2},\cdots,\pi_{q-1},\pi\}$.  Let
$f_{1}^{*}(x)=\frac{f(x)}{x+z^{*}}=\sum_{i=0}^{q-2}c_{i}x^{i}$,
$f_{2}^{*}(x)=\frac{f(x)(x+\pi)}{x+z^{*}}=\sum_{i=0}^{q-1}\tilde{c}_{i}x^{i}$ and
$f(x)=(x+z^{*})d(x)+\rho_{0}$ where $d(x)=\sum_{j=0}^{q-2}d_{j}x^{j}$.
Therefore,
\begin{equation*}%
Z^{*}=(g_{1}g_{2}^{w^{*}}g_{3}^{s^{*}})^{\frac{1}{x+z^{*}}}= 
g_{1}^{\frac{1}{x+{z^{*}}}}(g_{2}^{w^{*}}g_{3}^{s^{*}})^{\frac{1}{x+z^{*}}}.
\end{equation*} 

We have
\begin{equation*}%
\begin{split}%
&g_{1}^{\frac{1}{x+z^{*}}}=Z^{*}\cdot
(g_{2}^{w^{*}}g_{3}^{s^{*}})^{\frac{-1}{x+z^{*}}}\\
&=Z^{*}\cdot
(\bar{g}^{\frac{w^{*}((x+\pi)k-1)}{a}}\bar{g}^{\frac{\gamma_{2}
s^{*}((x+\pi)k-1)}{a}})^{\frac{-1}{x+z^{*}}}\\ 
&=Z^{*}\cdot \bar{g}^{\frac{-(w^{*}+ \gamma_{2}
s^{*})(x+\pi)k}{a(x+z^{*})}}\cdot \bar{g}^{\frac{w^{*}+ \gamma_{2}
s^{*}}{a(x+z^{*})}}\\ 
& =Z^{*}\cdot 
{g}^{\frac{-f(x)(w^{*}+\gamma_{2}
s^{*})(x+\pi)}{a(x+z^{*})}}\cdot g^{\frac{f(x)(w^{*}+\gamma_{2}
s^{*})}{a(x+z^{*})}}\\ 
&= Z^{*}\cdot {g}^{\frac{-(w^{*}+\gamma_{2}
s^{*})kf_{2}^{*}(x)}{a}}\cdot g^{\frac{(w^{*}+\gamma_{2}
s^{*})f_{1}^{*}(x)}{a}}\\ 
& =Z^{*}\cdot
\prod_{i=0}^{q-1}(g^{x^{i}})^{\frac{-\tilde{c}_{i}(w^{*}+\gamma_{2}
s^{*})k}{a}}\cdot 
\prod_{j=0}^{q-2}(g^{x^{j}})^{\frac{c_{j}(w^{*}+\gamma_{2}
s^{*})}{a}}.
\end{split}
\end{equation*}
		
Let $\Psi=Z^{*}\cdot
\prod_{i=0}^{q-1}(g^{x^{i}})^{\frac{-\tilde{c}_{i}(w^{*}+\gamma_{2}
s^{*})k}{a}}\cdot 
\prod_{j=0}^{q-2}(g^{x^{j}})^{\frac{c_{j}(w^{*}+\gamma_{2}
s^{*})}{a}}$.  We have 

\begin{equation*}%
\Psi=g_{1}^{\frac{1}{x+z^{*}}}=g^{\frac{f(x)}{x+z^{*}}}= 
g^{\frac{(x+z^{*})d(x)+\rho_{0}}{x+z^{*}}}=g^{d(x)}g^{\frac{\rho_{0}}{x+z^{*}}}. 
\end{equation*}%
Hence, 
\begin{equation*}%
\begin{split}
& g^{\frac{1}{x+z^{*}}}=(\Psi\cdot 
g^{-d(x)})^{\frac{1}{\rho_{0}}}=\Big(Z^{*}\cdot
\prod_{i=0}^{q-1}(g^{x^{i}})^{\frac{-\tilde{c}_{i}(w^{*}+\gamma_{2}
s^{*})k}{a}}\cdot \\
&
\prod_{j=0}^{q-2}(g^{x^{j}})^{\frac{c_{j}(w^{*}+\gamma_{2}
s^{*})}{a}}\cdot 
\prod_{k=0}^{q-2}(g^{x^{k}})^{-d_{k}}\Big)^{\frac{1}{\rho_{0}}}.
\end{split}
\end{equation*}
\medskip

\item{\sf Case-2.} $z^{*}\in\{\pi_{1},\pi_{2},\cdots,\pi_{q-1},\pi\}$. We have 
$z^{*}=\pi$ with the probability $\frac{1}{q}$. Since 
$\pi\notin\{\pi_{1},\pi_{2},\cdots,\pi_{q-1}\}$, $\mathcal{B}$ can output 
$g^{\frac{1}{x+\pi}}$ using the same technique above. 
\medskip

\item{\sf Case-3.} $z^{*}=\pi_{v}$, $Z^{*}=Z_{V}$, but 
$s^{*}\neq s_{v}$. Since 
$Z^{*}=(g_{1}g_{2}^{w^{*}}g_{3}^{s^{*}})^{\frac{1}{x+z^{*}}}$ 
and 
$Z_{V}=(g_{1}g_{2}^{w_{v}}g_{3}^{s_{v}})^{\frac{1}{x+\pi_{v}}}$. 
We have $g_{2}^{w^{*}}g_{3}^{s^{*}}=g_{2}^{w_{v}}g_{3}^{s_{v}}$, 
$g_{3}=g_{2}^{\frac{w^{*}-w_{v}}{s_{v}-s^{*}}}$ and 
$log_{g_{2}}g_{3}={\frac{w_{v}-w^{*}}{s^{*}-s_{v}}}$. $\mathcal{B}$ can 
use $\mathcal{A}$ to break the discrete logarithm assumption. Because JOC-$q$-SDH
assumption is included in discrete logarithm assumption, $\mathcal{B}$ can use $\mathcal{A}$ to break the JOC-$q$-SDH assumption. 
\end{itemize}
\medskip
	
Therefore, the advantage with which $\mathcal{B}$ can break the $q$-SDH 
assumption is 
\begin{equation*}%
\begin{split}%
&Adv_{\mathcal{B}}^{\mbox{JoC-q-SDH}}=\Pr[\mbox{{\sf Case-1}}]+\Pr[\mbox{{\sf 
Case-2}}]+\Pr[\mbox{{\sf Case-3}}]\\
& \geq (1-\frac{q}{p})\epsilon'(\ell)+\frac{q}{p}\times\frac{1}{q} 
\epsilon'(\ell)+\frac{1}{p}\times\frac{1}{p}\times
 (1-\frac{1}{p})\epsilon'(\ell)\\
& =(\frac{p-q}{p}+\frac{1}{p}+\frac{p-1}{p^{3}})\epsilon'(\ell).
\end{split}
\end{equation*}
\medskip
\end{IEEEproof}

\begin{theorem}\label{unlink}
Our scheme described in Fig. \ref{fig.setup}, Fig. \ref{fig.reg}, Fig.
\ref{fig.t-i}, Fig. \ref{fig.t-v}, Fig. \ref{fig.t-t}, Fig. \ref{fig.p-k-g} and
Fig. \ref{fig.p-t-v} is $\epsilon'(\ell)$-unlinkable if and only if the
$\epsilon(\ell)$-DaBDH assumption holds on the bilinear group
$(e,p,\mathbb{G}_{1},\mathbb{G}_{2},\mathbb{G}_{\tau})$, $H_{1}$,
$H_{2}$ and $H_{3}$ are secure cryptographic hash functions, and $H_{2}$ is a random
oracle,  where $\epsilon'(\ell)\geq\frac{\epsilon(\ell)}
{2\mathfrak{e}(1+q_{V{\mkern-6mu}A})}$, $\mathfrak{e}\approx 2.71$ is the
natural logarithm, $q_{V{\mkern-6mu}A}$ is the number of ticket verifiers
selected by $\mathcal{A}$ to query the {\sf Ticket-Verifier-Reg} oracle.
\end{theorem}

%

\begin{IEEEproof} If there exist a PPT adversary $\mathcal{A}$ that can break the unlinkability of our scheme, we can construct a PPT adversary which uses $\mathcal{A}$ as a subroutine to break the DaDBH assumption as follows. $\mathcal{C}$ flips an unbiased coin $\mu\in\{0,1\}$. If $\mu=0$,  $\mathcal{C}$ sends  $\mathbb{T}=\left(g,\mathfrak{g},g^{a},g^{b},g^{c},\mathfrak{g}^{b},\mathfrak{g}^{c}, \Upsilon=R\right)$ to $\mathcal{B}$ where $R\stackrel{R}{\leftarrow}\mathbb{Z}_{p}$; while if $\mu=1$,  $\mathcal{C}$ sends  $\mathbb{T}=\left(g,\mathfrak{g},g^{a},g^{b},g^{c},\mathfrak{g}^{b},\mathfrak{g}^{c}, \Upsilon=e(g,\mathfrak{g})^{abc}\right)$ to $\mathcal{B}$. $\mathcal{B}$ will outputs her guess $\mu'$ on $\mu$.
\medskip

\noindent{\sf Setup.} $\mathcal{B}$ selects $\alpha,\gamma,\gamma_{1},\gamma_{2},\gamma_{3},\gamma_{4},\gamma_{5}\stackrel{R}{\leftarrow}\mathbb{Z}_{p}$ and computes $\tilde{g}=g^{\gamma},$ $g_{1}=g^{\gamma_{1}},g_{2}=g^{\gamma_{2}},g_{3}=g^{\gamma_{3}},$ $\vartheta_{1}=g^{\gamma_{4}}$, $\vartheta_{2}=g^{\gamma_{5}}$,  $Y_{A}=\mathfrak{g}^{\alpha}$ and $\tilde{Y}_{A}=(g^{b})^{\gamma}$. $\mathcal{B}$ selects $H_{1}:\{0,1\}^{*}\rightarrow\mathbb{Z}_{p}$, $H_{2}:\{0,1\}\rightarrow \mathbb{G}_{2}$ and $H_{3}:\{0,1\}^{*}\rightarrow \{0,1,\}^{\ell'}~(\ell'\leq\ell)$. $\mathcal{B}$ sends $PP=(e,p,\mathbb{G}_{1},\mathbb{G}_{2},\mathbb{G}_{\tau},\tilde{g},g,g_{1},g_{2},g_{3},\mathfrak{g},\vartheta_{1},\vartheta_{2},Y_{A},\tilde{Y}_{A},H_{1},H_{2},H_{3})$ to $\mathcal{A}$. We imply that the master secret key is $MSK=(\alpha,\beta=b)$.
\medskip

\noindent{\sf $H_{2}$-queries.} $\mathcal{B}$ maintains a list $H_{2}^{list}$ which consists of the tuples $(ID_{i},\Psi_{i},\gamma_{i},coin)$ and is initially empty. When $\mathcal{A}$ queries $H_{2}$ with an identity $ID_{j}$, $\mathcal{B}$ works as follows:
\begin{enumerate}
\item If the query $ID_{j}$ is  in the $H_{2}^{list}$,  $\mathbb{B}$ returns $H_{2}(ID_{j})=\Psi_{j}\in\mathbb{G}_{2}$;

\item If the query $ID_{j}$ is not in the $H_{2}^{list}$, $\mathcal{B}$ flips a random coin $coin\in\{0,1\}$ so that $\Pr[coin=0]=\delta$ from some $\delta$ which will be determine latter.  $\mathcal{B}$ selects $\gamma_{j}\stackrel{R}{\leftarrow}\mathbb{Z}_{p}$. If $coin=0$,  $\mathcal{B}$ computes $\Psi_{j}=\mathfrak{g}^{\gamma_{j}}$; If $coin=1$,  $\mathcal{B}$ computes $\Psi_{j}=(\mathfrak{g}^{c})^{\gamma_{j}}$. $\mathcal{B}$ adds $(ID_{j},\Psi_{j},\gamma_{j},coin)$ into $H_{2}^{list}$, and returns $H_{2}(ID_{j})=\Psi_{j}$ to $\mathcal{A}$.
\end{enumerate}
\medskip

\noindent{\bf Phase 1.} $\mathcal{A}$ can adaptively make the following queries.
\medskip

\noindent{\sf Registration Query.} $\mathcal{A}$ can make the following  registration queries.
 \begin{enumerate}
 \item{\em Ticket Issuer Registration Query.}  $\mathcal{B}$ selects $x_{i}\stackrel{R}{\leftarrow}\mathbb{Z}_{p}$ and computes $Y_{I}=g^{x^{i}}$ and $\tilde{Y}_{I}=\mathfrak{g}^{x_{i}}$. $\mathcal{B}$ selects $d_{i},e_{i}\stackrel{R}{\leftarrow}\mathbb{Z}_{p}$ and computes $\sigma_{I}=(g_{1}g_{2}^{d_{i}}Y_{I})^{\frac{1}{\alpha+e_{i}}}$. $\mathcal{B}$ sends $(ID_{I},Y_{I},\tilde{Y}_{I},\sigma_{I})$ to $\mathcal{A}$.

  \item{\em Ticket Verifier Registration Query.} Let $Corrupt_{V}$ be the set consisting of the identities of ticket verifiers corrupted by $\mathcal{A}$. $\mathcal{A}$  submits a verifier' identity  $ID_{V}$. $\mathcal{B}$ checks the $H_{2}^{list}$ and obtains the tuple $(ID_{V},\Psi_{V},\gamma_{v},coin)$. $\mathcal{B}$ works as follows: (1) if  $coin=1$,  $\mathcal{B}$ aborts; (2) if $ID_{V}\in Corrupt_{V}$ and $coin=0$, $\mathcal{B}$ selects $d_{v},e_{v}\stackrel{R}{\leftarrow}\mathbb{Z}_{p}$ and  computes $SK_{V}=(\mathfrak{g}^{b})^{\gamma_{v}}=(\mathfrak{g}^{\gamma_{v}})^{b}=\Psi_{V}^{b}=H_{2}(ID_{V})^{b}$ and $\sigma_{V}=(g_{1}g_{2}^{d_{v}}G^{H_{1}(ID_{V})})^{\frac{1}{\alpha+e_{v}}}$. $\mathcal{B}$ and sends $(SK_{V},\sigma_{V})$ to $\mathcal{A}$;  (3) if $ID_{V}\notin Corrupt_{V}$ and $coin=0$,  $\mathcal{B}$ selects $d_{v},e_{v}\stackrel{R}{\leftarrow}\mathbb{Z}_{p}$,  and  computes $SK_{V}=(\mathfrak{g}^{b})^{\gamma_{v}}=(\mathfrak{g}^{\gamma_{v}})^{b}=\Psi_{V}^{b}=H_{2}(ID_{V})^{b}$ and  $\sigma_{V}=(g_{1}g_{2}^{d_{v}}G^{H_{1}(ID_{V})})^{\frac{1}{\alpha+e_{v}}}$. $\mathcal{B}$ sends $\sigma_{V}$ to $\mathcal{A}$.  Let $VK$ be a set consisting of the ticket verifier registration query information $(ID_{V},SK_{V},\sigma_{V})$.

 \item{\em User Registration Query.} When $\mathcal{A}$ submits a user's identity $ID_{U}$ with public key $PK_{U}$, $\mathcal{B}$ selects $d_{u},e_{u}\stackrel{R}{\leftarrow}\mathbb{Z}_{p}$ and computes $\sigma_{U}=(g_{1}g_{2}^{d_{u}}Y_{U})^{\frac{1}{\alpha+e_{u}}}$. $\mathcal{B}$ sends $\sigma_{U}$ to $\mathcal{A}$.
 \item{\em Central Verifier Registration Query.} $\mathcal{B}$ selects $x_{cv},d_{cv},e_{cv}\stackrel{R}{\leftarrow}\mathbb{Z}_{p}$ and computes $Y_{CV}=g^{x_{cv}}$ and $\sigma_{CV}=(g_{1}g_{2}^{d_{cv}}Y_{CV})^{\frac{1}{\alpha+e_{cv}}}$. $\mathcal{B}$ sends $(Y_{CV},\sigma_{CV})$ to $\mathcal{A}$.
  \end{enumerate}
\medskip

\noindent{\sf Ticket Issuing Query.} $\mathcal{A}$ submits a set of  service information $J_{U}$ consisting of ticket verifiers, a set of pseudonyms $\{(P{V},Q_{V})|ID_{V}\in J_{U}\}$ and a proof $\prod_{U}^{1}$ of the credential $\sigma_{U}$ and the pseudonyms. $\mathcal{B}$ verifies $\prod_{U}^{1}$ and $e(\bar{\sigma}_{U},Y_{U})\stackrel{?}{=}e(\tilde{\sigma_{U}},\mathfrak{g})$. If there are correct, $\mathcal{B}$ selects $r_{u}\stackrel{R}{\leftarrow}\mathbb{Z}_{p}$ and computes $R_{U}=g^{r_{u}}$. For $ID_{V}\in J_{U}$, $\mathcal{B}$ selects $t_{v},w_{v},z_{v}\stackrel{R}{\leftarrow}\mathbb{Z}_{p}$, and computes $D_{V}=H_{3}(R_{U}||ID_{V}),E_{V}^{1}=e(\tilde{Y}_{A},H_{2}(ID_{V}))^{t_{v}},E_{V}^{2}=\tilde{g}^{t_{v}},E_{V}^{3}=(\vartheta_{1}\vartheta_{2}^{H_{1}(TP||Text_{1})})^{t_{v}}, K_{V}=\tilde{g}^{H_{1}(ID_{V})}Y_{CV}^{t_{v}}$, $s_{v}=H_{1}(P_{V}||Q_{V}||E_{V}^{1}||E_{V}^{2}||E_{V}^{3}||K_{V}||text_{2})$ and $Z_{V}=(g_{1}g_{2}^{w_{v}}g_{3}^{s_{v}})^{\frac{1}{x_{i}+z_{v}}}$. The authentication tage is $Tag_{V}=((P_{V},Q_{V}),(E_{V}^{1},E_{V}^{2},E_{V}^{3},K_{V},Text_{1},Text_{2}),(s_{v},w_{v},z_{v},$ $Z_{v}))$. For the central verifier $ID_{CV}\in J_{U}$, $\mathcal{B}$ selects $w_{cv},z_{cv}\stackrel{R}{\leftarrow}\mathbb{Z}_{p}$, and computes $s_{cv}=H_{1}(s_{1}||s_{2}||\cdots||s_{|J_{U}|})$ and $Z_{CV}=(g_{1}g_{2}^{w_{cv}}g_{3}^{s_{cv}})^{\frac{1}{x_{i}+z_{cv}}}$. The ticket is $T_{U}=\{(D_{V},Tag_{V})|ID_{V}\in J_{U}\}\cup\{s_{cv},w_{cv},z_{cv},Z_{CV}\}$. $\mathcal{B}$ sends $(R_{U},T_{U})$ to $\mathcal{A}$.
\medskip

\noindent{\sf Ticket Trace Query.} $\mathcal{A}$ adaptively submits a ticket $T_{U}$. Let $\Omega_{U}=\{\}$. For each $Tag_V$ in $T_U$, $\mathcal{B}$ works as follows:
  (1)   a) Compute: $Y_{U}=\frac{P_{V}}{Q_{V}^{x_{cv}}}$ and $g^{H_{1}(ID_{V})}=\frac{K_{V}}{(E_{V}^{3})^{x_{cv}}}$; b) Look up $g^{H_{1}(ID_V)}$ and $\mathcal{V}$'s identity.
$\mathcal{B}$ check: 
 (c1) $s_{v}\stackrel{?}{=}H_{1}(P_{V}||Q_{V}||E_{V}^{1}||E_{V}^{2}||E_{V}^{3}||K_{V}||Text_{1})$;
 (c2) $e(Z_{V},Y_{I}\mathfrak{g}^{z_{v}})\stackrel{?}{=}e(g_{1}g_{2}^{w_{v}}g_{3}^{s_{v}},\mathfrak{g})$;
(d) If (c1) and (c2) hold, set $\Omega_U=\Omega_U\cup\{ID_{V}\}$; otherwise abort. 
 (e) Verify $Y_U$ remains the same for all tags.   (2) $s_{cv}\stackrel{?}{=}H_{1}(s_{1}||s_{2}||\cdots||s_{|J_{U}|})$;  (3) $e(Z_{CV},\tilde{Y}_{I}\mathfrak{g}^{z_{cv}})\stackrel{?}{=}e(g_{1}g_{2}^{w_{cv}}g_{3}^{s_{cv}},\mathfrak{g})$.
   If (1), (2) and (3) hold, $\mathcal{CV}$ can determine that the service    
  information of $\mathcal{U}$ with public key $Y_{U}$ is:  $J_{U}=\Omega_U$;
  otherwise, the trace has failed.
\medskip

\noindent{\sf Proxy Key Generation Query.} $\mathcal{A}$ adaptively submits two identities $ID_{V}$ and $ID_{V'}$. $\mathcal{B}$ chekcs $H_{2}^{list}$ and obtains $(ID_{V},\Psi_{V},\gamma_{v},coin)$ and  $(ID_{V'},\Psi_{V'},\gamma'_{v},coin)$. If any $coin=1$, $\mathcal{B}$ aborts; otherwise, $\mathcal{B}$ checks $VK$, and obtains $(ID_{V},SK_{V},\sigma_{V})$ and $(ID_{V'},SK_{V'},\sigma_{V'})$. $\mathcal{B}$ selects $\beta_{v}\stackrel{R}{\leftarrow}\mathbb{Z}_{p}$ and computes $RK_{1}=\tilde{g}^{\beta_{v}}$ and $RK_{2}=(\vartheta_{1}\vartheta_{2}^{H_{1}(TP||Text_{1})})^{\beta_{v}}\cdot \frac{SK_{V}}{SK_{V'}}$. $\mathcal{B}$ responds $\mathcal{A}$ with $RK_{\mathcal{V}\rightarrow\mathcal{V}}=(RK_{1},RK_{2})$. Let $PQ$ be a set consisting of the proxy key generation query. $\mathcal{C}$ adds $(ID_{V},ID_{V'},TP,RK_{\mathcal{V}\rightarrow\mathcal{V}'})$ into $PQ$.
\medskip

\noindent{\sf Proxy Ticket Validation Query.} $\mathcal{A}$ adaptively submits $(\prod_{V}^{2},Tag_{V},ID_{V},ID_{V'})$. $\mathcal{B}$ checks wether  $(ID_{V},ID_{V'},TP,RK_{\mathcal{V}\rightarrow\mathcal{V}'})$ is in $PQ$. If it is not, $\mathcal{B}$ aborts; otherwise, $\mathcal{B}$ works as follows: If  $(e_{v},w_{v},s_{v},Z_{V})\in T_{V'}$, aborts; otherwise, adds $(e_{v},w_{v},s_{v},Z_{V})$ in $T_{V'}$ and goes to the next step. Computes:   $\Theta_{1}=RK_{2}\cdot SK_{V'}$ and 
 $\Theta_{2}=\frac{e(E_{V}^{2},\Theta_{1})}{e(RK_{1},E_{V}^{3},)}$.
 Checks:
 (1) The correctness of $\prod_{U}^{2}$;
(2)  $s_{v}\stackrel{?}{=}H_{1}(P_{V}||Q_{V}||E_{V}^{1}||E_{V}^{2}||E_{V}^{3}||K_{V}||Text_{2})$;
 (3) $\Theta_{2}\stackrel{?}{=}E_{V}^{1}$;
(4)  $e(Z_{V},Y_{I}\mathfrak{g}^{z_{v}})\stackrel{?}{=}e(g_{1}g_{2}^{w_{v}}g_{3}^{s_{v}},\mathfrak{g})$;
 If (1), (2), (3) and (4)  hold, $\mathcal{B}$ sends $(1,Tag_{V})$ to $\mathcal{A}$ to indicate success;  otherwise, $(0,Tag_{V})$ is returned to $\mathcal{A}$ to indicate failure.
\medskip

\noindent{\bf Challenge.}  $\mathcal{A}$ submits two verifiers $((P_{V_{0}}^{*},Q_{V_{0}}^{*}),ID_{V_{0}^{*}})$ and $((P_{V_{1}}^{*},Q_{V_{1}}^{*}),ID_{V_{1}^{*}})$ with the limitation that $ID_{V_{0}^{*}},ID_{V_{1}^{*}}\notin Corrupt_{V}$ and  $(ID_{V_{0}},ID_{V_{1}},TP,RK_{\mathcal{V}_{0}^{*}\rightarrow\mathcal{V}_{1}^{*}})\notin PQ$. $\mathcal{C}$ flips an unbiased coin with $\{0,1\}$ and obtains a bit $\varrho\in\{0,1\}$.  $\mathcal{C}$ sets $J_{U^{*}}=\{ID_{V_{\varrho}^{*}}\}$. $\mathcal{B}$ runs the algorithm for  the $H_{2}^{list}$ and obtains $(ID_{V_{\varrho}^{*}},\Psi^{*},\gamma^{*},coin)$. If $coin=0$, $\mathcal{B}$ aborts; otherwise, $H_{2}(ID_{V_{\varrho}^{*}})=\Psi^{*}=(\mathfrak{g}^{c})^{\gamma^{*}}$. $\mathcal{B}$ selects $r^{*},w^{*},z^{*},w_{cv},z_{cv}\stackrel{R}{\leftarrow}\mathbb{Z}_{p}$, and computes $R^{*}=\bar{g}^{r^{*}}$, $D_{V^{*}}=H_{3}(R^{*}||ID_{V_{\varrho}^{*}})$, $E_{V_{\varrho}^{*}}^{1}=\Upsilon$, $E_{V_{\varrho}^{*}}^{2}=(g^{a})^{\gamma}=\tilde{g}^{a}$, $E_{V_{\varrho}^{*}}^{3}=(g^{a})^{\gamma_{4}}(g^{a})^{\gamma_{5}H_{1}(TP||Text_{1})}=(\vartheta_{1}\vartheta_{2}^{H_{1}(TP||Text_{1})})^{a}$, $K_{V_{\varrho}^{*}}=g^{H_{1}(ID_{V_{\varrho}^{*}})}(g^{a})^{x_{cv}}=g^{H_{1}(ID_{V_{\varrho}^{*}})}Y_{CV}^{a}$, $s^{*}=H_{1}(P_{V_{\varrho}^{*}}||Q_{V_{\varrho}^{*}}||E_{V_{\varrho}^{*}}^{1}||E_{V_{\varrho}^{*}}^{2}||E_{V_{\varrho}^{*}}^{3}||K_{V_{\varrho}^{*}}||Text_{2})$, $Z_{V_{\varrho}^{*}}=(g^{\gamma_{1}}g^{\gamma_{2}w^{*}}g^{\gamma_{3}s^{*}})^{\frac{1}{x_{i}+z^{*}}}=(g_{1}g_{2}^{w^{*}}g_{3}^{s^{*}})^{\frac{1}{x_{i}+z^{*}}}$, $s_{cv}=H_{1}(s^{*})$, $Z_{CV}=(g^{\gamma_{1}}g^{\gamma_{2}w_{cv}}g^{\gamma_{3}s_{cv}})^{\frac{1}{x_{i}+z_{cv}}}=(g_{1}g_{2}^{w_{cv}}g_{3}^{s_{cv}})^{\frac{1}{x_{i}+z_{cv}}}$. The challenged ticket is $Tag_{V_{\varrho}^{*}}=((P_{V_{\varrho}^{*}},Q_{V_{\varrho}^{*}}),(E_{V_{\varrho}^{*}}^{1},E_{V_{\varrho}^{*}}^{2},E_{V_{\varrho}^{*}}^{3},T_{V_{\varrho}^{*}},Text_{2}),(s^{*},w^{*},z^{*},Z_{V_{\varrho}^{*}}))$.

Let $T_{U}^{*}=(Tag_{V_{\varrho}^{*}})\cup (s_{cv},w_{cv},z_{cv},Z_{CV})$. $\mathcal{B}$ sends $(D_{V_{\varrho}^{*}},T_{U}^{*})$ to $\mathcal{A}$.
\medskip

\noindent{\bf Phase 2.} This is the same as in {\bf Phase 1.}
\medskip

\noindent{\bf Output.} $\mathcal{A}$ outputs his guess $\varrho'$ on $\varrho$. If $\varrho'=\varrho$, $\mathcal{B}$ outputs $\mu'=1$; otherwise, $\mathcal{B}$ outputs $\mu'=0$.
\medskip

Now, we compute the probability with which $\mathcal{B}$ does not abort. Suppose $\mathcal{A}$ totally makes $q_{V}$ ticket verification registration queries. Hence, the probability that $\mathcal{B}$ does not aborts in {\bf Phase 1} and {\bf Phase 2} is $\delta^{q_{v}}$ and the probability that $\mathcal{B}$ does not aborts in {\bf Challenge} is $(1-\delta)$. The probability that $\mathcal{B}$ does not aborts in the game is $\delta^{q_{V}}(1-\delta)$ and achieves the maximum value $\frac{1}{\mathfrak{e}(1+q_{V})}$ when $\delta=\frac{q_{V}}{1+q_{V}}$ ( $\mathfrak{e}\approx 2.71$ is the natural logarithm). Therefore, $\mathcal{B}$ does not abort the game with probability at least $\frac{1}{\mathfrak{e}(1+q_{V})}$.
\medskip

If $\mu=1$, $Tag_{V_{\varrho}^{*}}$ and $T_{U}^{*}$ are valid. Hence, $\mathcal{A}$ outputs $\varrho'=\varrho$ with probability at least $\frac{1}{2}+\epsilon'(\lambda)$. Since $\varrho'=\varrho$, $\mathcal{B}$ outputs $\mu'=1$. We have $\Pr[\mu'=\mu|\mu=1]\geq \frac{1}{2}+\epsilon'(\lambda)$. If $\mu=0$, $Tag_{V_{\varrho}^{*}}$ and $T_{U}^{*}$ are invalid. Hence, $\mathcal{A}$ outputs $\varrho'\neq\varrho$ with probability  $\frac{1}{2}$. Since $\varrho'\neq\varrho$, $\mathcal{B}$ outputs $\mu'=0$. We have $\Pr[\mu'=\mu|\mu=0]=\frac{1}{2}$. The advantage with which $\mathcal{B}$ outputs $\mu'=\mu$ is 
$\left|\frac{1}{2}\Pr[\mu'=\mu|\mu=1]-\frac{1}{2}\Pr[\mu'=\mu|\mu=0]\right|
 \geq\frac{1}{2}\Pr[\mu'=\mu|\mu=1] - \frac{1}{2}\Pr[\mu'=\mu|\mu=0]
= \frac{1}{2}\times(\frac{1}{2}+\epsilon'(\lambda))-\frac{1}{2}\times \frac{1}{2}=\frac{1}{4}+\frac{\epsilon'(\lambda)}{2}-\frac{1}{4}
= \frac{\epsilon'(\lambda)}{2}$.

Therefore, the total advantage with which $\mathcal{B}$ can break the DaDBH assumption is at least $\frac{1}{\mathfrak{e}(1+q_{V})}\times\frac{\epsilon'(\lambda)}{2}=\frac{\epsilon'(\lambda)}{2\mathfrak{e}(1+q_{V})}$.
\end{IEEEproof}
\medskip

\begin{theorem}\label{theorem:trace}%
Our scheme described in Fig. \ref{fig.setup}, Fig. \ref{fig.reg}, Fig. \ref{fig.t-i},
Fig. \ref{fig.t-v}, Fig.  \ref{fig.t-t}, Fig. \ref{fig.p-k-g} and Fig.
\ref{fig.p-t-v} is $(\varrho,\epsilon(\ell))$-traceable if the JoC-$q$-SDH
assumption holds on the bilinear group
$(e,p,\mathbb{G}_{1},\mathbb{G}_{2},\mathbb{G}_{\tau})$ with the advantage at
most  $\epsilon_{1}(\ell)$, the DL assumption holds on the group
$\mathbb{G}_{1}$ with the advantage at most $\epsilon_{2}(\ell)$, and $H_{1},
H_{2}$ and $H_{3}$ are  secure cryptographic hash functions, where
$\epsilon(\ell)=max\left\{\frac{\epsilon_{1}(\ell)}{2}(\frac{p-q}{p}+
\frac{1}{p}+\frac{p-1}{p^{3}}),\frac{\epsilon_{2}(\ell)}{2}\right\}$, $\varrho$
is the total number of ticket issuing queries made by $\mathcal{A}$ and
$\varrho<q$.%
\end{theorem}


\begin{IEEEproof}
Suppose that there exists an adversary $\mathcal{A}$ that can break the traceability of our scheme, we can construct an algorithm $\mathcal{B}$ which can use $\mathcal{A}$ as a subroutine to break the JoC-$q$-SDH assumption or DL assumption. Given  a $(q+3)$-tuple 
$(g,g^{x},\cdots,g^{x^{q}},\mathfrak{g},
\mathfrak{g}^{x})\in\mathbb{G}_{1}^{q+1}\times\mathbb{G}_{2}^{2}$, $\mathcal{B}$ will output $(c,g^{\frac{1}{x+c}})\in\mathbb{Z}_{p}\times\mathbb{G}_{1}$ where $c\neq x$. 
\medskip

\noindent{\sf Setup.} $\mathcal{B}$ selects $\pi_{1},\pi_{2},\cdots,\pi_{q-1}\stackrel{R}{\leftarrow}\mathbb{Z}_{p}$, and sets $f(x)=\prod_{i=1}^{q-1}(x+\pi_{i})=\sum_{i=0}^{q-1}\theta_{i}x^{i}$, $f_{i}(x)=\frac{f(x)}{x+\pi_{i}}=\sum_{j=0}^{q-2}\omega_{i_{j}}x^{j}$, $\tilde{g}=\prod_{i=0}^{q-1}(g^{x^{i}})^{\chi_{i}}=g^{f(x)}$, $\hat{g}=\prod_{i=0}^{q-1}(g^{x^{i+1}})^{\theta_{i}}=\tilde{g}^{x}$.  $\mathcal{B}$ selects $\pi,a,k\stackrel{R}{\leftarrow}\mathbb{Z}_{p}$ and computs $g_{2}=((\hat{g}\bar{g}^{\pi})^{k}\bar{g}^{-1})^{\frac{1}{a}}=\bar{g}^{\frac{(x+\pi)k-1}{a}}$.  $\mathcal{B}$ 
 selects  $\gamma_{1},\gamma_{2},\gamma_{3}\stackrel{R}{\leftarrow}\mathbb{Z}_{p}$  and  computes $g_{1}=\bar{g}^{\gamma_{1}}$, $g_{3}=g_{2}^{\gamma_{2}}$ and $\bar{g}=g^{\gamma_{3}}$. $\mathcal{B}$ selects hash functions $H_{1}:\{0,1\}^{*}\rightarrow\mathbb{Z}_{p}$, $H_{2}:\{0,1\}\rightarrow \mathbb{G}_{2}$ and $H_{3}:\{0,1\}^{*}\rightarrow \{0,1\}^{\ell'}~(\ell'\leq\ell)$. $\mathcal{B}$ selects $\alpha,\beta,\stackrel{R}{\leftarrow}\mathbb{Z}_{p}$ and $\vartheta_{1},\vartheta_{2}\stackrel{R}{\leftarrow}\mathbb{G}_{2}$. $\mathcal{B}$ computes ${Y}_{A}=\mathfrak{g}^{\alpha}$ and  $\tilde{Y}_{A}=\tilde{g}^{\beta}$. The master secret key is $MSK=(\alpha,\beta)$ and the public parameters are $PP=(e,p,\mathbb{G}_{1},\mathbb{G}_{2},\mathbb{G}_{\tau},\tilde{g},\bar{g},g_{1},g_{2},g_{3},\mathfrak{g},\vartheta_{1},\vartheta_{2},Y_{A},\tilde{Y}_{A},$ $H_{1},H_{2},H_{3})$.  
\medskip

\noindent{\sf Registration Query.} $\mathcal{A}$ adaptively makes the following queries:
\begin{enumerate}
\item{\em Ticket Issuer Registration Query.} $\mathcal{B}$ sets $Y_{I}=\tilde{g}^{x}$ and $\tilde{Y}_{I}=\mathfrak{g}^{x}$. $\mathcal{B}$ selects $d_{i},e_{i}\stackrel{R}{\leftarrow}\mathbb{Z}_{p}$ and computes $\sigma_{I}=(g_{1}g_{2}^{d_{i}}Y_{I})^{\frac{1}{\alpha+e_{i}}}$. $\mathcal{B}$ sends $(\sigma_{I},Y_{I},\tilde{Y}_{I})$ to $\mathcal{A}$.

\item{\em Ticket Verifier Registration Query.} $\mathcal{A}$ submits an identity $ID_{V}\in\{0,1\}^{*}$. $\mathcal{B}$ selects $d_{v},e_{v}\stackrel{R}{\leftarrow}\mathbb{Z}_{p}$, and computes $\sigma_{V}=(g_{1}g_{2}^{d_{v}}\tilde{g}^{H_{1}(ID_{V})})^{\frac{1}{\alpha+e_{v}}}$ and $SK_{V}=H_{2}(ID_{V})^{\beta}$. $\mathcal{B}$ sends $(\sigma_{V},SK_{V})$ to $\mathcal{A}$.

\item{\em User Registration Query.} $\mathcal{A}$ submits an identity $ID_{U}$ and the corresponding public key $Y_{U}$. $\mathcal{C}$ selects $e_{u},d_{u}\stackrel{R}{\leftarrow}\mathbb{Z}_{p}$ and computes $\sigma_{U}=(g_{1}g_{2}^{d_{u}}Y_{U})^{\frac{1}{\alpha+e_{u}}}$. $\mathcal{B}$ returns $\sigma_{U}$ to $\mathcal{A}$.

\item{\em Central Verifier Registration Query.} $\mathcal{A}$ submits an identity $ID_{CV}\in\{0,1\}^{*}$ and the corresponding public key $Y_{CV}$. $\mathcal{B}$ selects $d_{cv},e_{cv}\stackrel{R}{\leftarrow}\mathbb{Z}_{p}$ and computes $\sigma_{CV}=(g_{1}g_{2}^{d_{v}}Y_{CV})^{\frac{1}{\alpha+e_{cv}}}$. $\mathcal{B}$ returns $\sigma_{CV}$ to $\mathcal{A}$.
\end{enumerate}

\medskip

\noindent{\sf Ticket Issuing Query.}  $\mathcal{A}$ adaptively submits  a set of service information $J_{U}$ and a set of pseudonyms $(P_{V},Q_{V})$ for $ID_{V}\in J_{U}$ and a proof $\prod_{U}^{1}: \mbox{PoK}\LARGE\{(x_{u},d_{u},e_{u},\sigma_{U},y,y_{1},y_{2},$ $y_{4},(k_{v})_{{V}\in J_{U}}): \frac{\tilde{\sigma}_{U}}{\bar{A}_{U}}=\bar{\sigma}_{U}^{-e_{u}}g_{2}^{y_{2}} 
 \wedge~ g_{1}^{-1}= \bar{A}_{U}^{-y_{4}}g_{2}^{y}\tilde{g}^{x_{u}}\wedge~(P_{V}= \tilde{g}^{x_{u}}{Y}_{CV}^{k_{v}} \wedge Q_{V}=\tilde{g}^{k_{v}})_{{V}\in J_{U}}\LARGE\}$. $\mathcal{B}$ checks the proof $\prod_{U}^{1}$ and  $e(\bar{\sigma}_{U},Y_{U})\stackrel{?}{=}e(\tilde{\sigma}_{U},\mathfrak{g})$. If each of them is incorrect, $\mathcal{B}$ aborts; otherwise, $\mathcal{B}$ goes to the next step. $\mathcal{B}$ chooses $r_{u}\stackrel{R}{\leftarrow}\mathbb{Z}_{p}$, and computes $R_{u}=\bar{g}^{r_{u}}$.  For $ID_{V}\in J_{U}$ and let $f_{v}(x)=\frac{f(x)}{x+\pi_{v}}=\sum_{j=0}^{q-2}\omega_{v_{j}}x^{j}$, $\mathcal{B}$ selects $t_{v},w_{v},z_{v}\stackrel{R}{\leftarrow}\mathbb{Z}_{p}$, and computes $D_{V}=H_{3}(R_{U}||ID_{V})$, $E_{V}^{1}=e(\tilde{Y}_{A},H_{2}(ID_{V}))^{t_{v}}$, $E_{V}^{2}=\tilde{g}^{t_{v}}$, $E_{V}^{3}=(\vartheta_{1}\vartheta_{2}^{H_{1}(TP||Text_{1})})^{t_{v}}$, $K_{V}=\tilde{g}^{h_{1}(ID_{V})}Y_{CV}^{t_{v}}$, $s_{v}=H_{1}(P_{V}||Q_{V}||E_{V}^{1}||E_{V}^{2}||E_{V}^{3}||K_{V}||Test_{1}||Test_{2})$ and $Z_{V}=\prod_{j=0}^{q-2}(g^{x^{j}})^{\omega_{v_{j}}(\gamma_{1}+\frac{(\pi k-1)(w_{v}+\gamma_{2}s_{v})}{a})}\prod_{j=0}^{q-2}(g^{x^{j+1}})^{\frac{\omega_{v_{j}}k (w_{v}+\gamma_{2}s_{v})}{a}}$.
 $(w_{v},\pi_{v},Z_{V})$ is a valid BBS+ signature on $s_{v}$ (see the proof in Section  \ref{prooftheorem:3}).

Let $f_{cv}=\frac{f(x)}{x+\pi_{cv}}=\sum_{j=0}^{q-2}\omega_{c_{j}}x^{j}$ with $\pi_{cv}\in\{\pi_{1},\pi_{2},\cdots,\pi_{q-1}\}$. $\mathcal{B}$ selects $w_{cv}\stackrel{R}{\leftarrow}\mathbb{Z}_{p}$ and computes $s_{cv}=H_{1}(s_{1}||s_{2}||\cdots||s_{|J_{U}|})$ and $Z_{CV}=\prod_{j=0}^{q-2}(g^{x^{j}})^{\omega_{c_{j}}(\gamma_{1}+\frac{(\pi k-1)(w_{cv}+\gamma_{2}s_{cv})}{a})}\prod_{j=0}^{q-2}(g^{x^{j+1}})^{\frac{\omega_{c_{j}}k (w_{cv}+\gamma_{2}s_{cv})}{a}}$.

When the $q$-th signature is required, $\mathcal{B}$ sets $w_{cv}=a\gamma_{1}-s_{cv}\gamma_{2}$ and $Z_{CV}=\bar{g}^{\gamma_{1}k}$.  $(w_{cv},\pi,Z_{CV})$ is a valid BBS+ signature (see the proof of {\em Theorem \ref{theorem:unforg}}).
The ticket is $T_{U}=\{(D_{V},Tag_{V})|ID_{V}\in J_{U}\}\cup\{s_{cv},w_{cw},$ $\pi_{cv},Z_{CV}\}$. Let $TQ$ be a set consisting of the ticket information queried by $\mathcal{A}$ and initially empty. $\mathcal{B}$ adds $(R_{U},T_{U},Text_{1},Text_{2})$ into $TQ$ and sends it to $\mathcal{A}$.
\medskip

\noindent{\sf Output.} $\mathcal{A}$ outputs a ticket $T_{U}^{*}=\{(D_{V}*,Tag_{V^{*}})|ID_{V^{*}}\in J_{U}\}\cup\{s_{cv}^{*},w_{cv}^{*},z_{cv}^{*},Z_{CV}^{*}\}$. If more than two users' public keys are included in $T_{U}^{*}$, $T_{U}^{*}$ is generated incorrectly and $\mathcal{B}$ aborts; otherwise, we consider the following two types of forgers. 
Type-1 forger outputs a ticket $T_{U}^{*}$  
which includes at least a new pseudonym $ (\hat{P}_{V},\hat{Q}_{V})$  which is not 
included in any ticket queried by $\mathcal{A}$. Type-2 forger outputs a 
ticket $T_{U}^{*}$ which includes the same pseudonyms  included in a ticket 
$T_{U}\in TQ$ queried by $\mathcal{A}$, but can be trace to a user 
$\mathcal{U}'$ whose secrete key $x'$ is not known by $\mathcal{A}$. Let 
$(x_{u'},Y_{U'})$ be the secret-public key pair of $\mathcal{U}'$.

\begin{itemize}

\item{\em Type-1:} If there is a pseudonym $(\hat{P}_{V},\hat{Q}_{V})\subset Tag'_{V}\in 
T_{U^{*}}$ and $(P'_{V},Q'_{V})\notin TQ$.  Let $\widehat{Tag}_{V}=((\hat{P}_{V},\hat{Q}_{V}),(\hat{E}^{1}_{V},\hat{E}^{2}_{V},\hat{E}^{3}_{V},\hat{K}_{V},Text_{1},Text_{2}),(\hat{s}_{v},\hat{w}_{v},$ $\hat{z}_{v},\hat{Z}_{V}))$. $\mathcal{A}$  forged a 
signature $(\hat{w}_{v},\hat{z}_{v},\hat{Z}_{V})$ on $\hat{s}_{v}$ where 
$\hat{s}_{v}=H_{1}(\hat{P}_{V}||\hat{Q}_{V}||$ $\hat{E}^{1}_{V}||\hat{E}^{2}_{V}||\hat{E}^{3}_{V}||\hat{K}_{V}||Text_{1}||Text_{2})$. Hence, 
$\mathcal{B}$ can use $\mathcal{A}$ to break the JOC-$q$-SDH assumption 
by using the technique in the proof in Section \ref{prooftheorem:3}.

\item{\em Type-II.} If all pseudonyms $(P_{V},Q_{V})\subset Tag_{V}\in 
T_{U}^{*}$ and $(P_{V},Q_{V})\in TQ$. In $T_{U}^{*}$, there is a pseudonym 
$(P_{CV},Q_{CV})$ generated for the central verifier. If $\mathcal{A}$ can 
generate a proof   $\prod_{U}^{1}: \mbox{PoK}\{(x_{u}',z_{cv}): 
P_{CV}=\xi^{x_{u}'}{Y}_{CV}^{k_{cv}}~\wedge~Q_{V}=\tilde{g}^{k_{cv}}\}$, $\mathcal{B}$ 
can use the rewinding technique to extract the knowledge of $(x_{u}',k_{cv})$ from 
$\mathcal{A}$, namely given $(\tilde{g},Y_{U'})$, $\mathcal{B}$ can output a $x_{u'}$ such 
that $Y_{U'}=\tilde{g}^{x_{u'}}$. Hence, $\mathcal{B}$ can use $\mathcal{A}$ to break the 
discrete logarithm assumption.

\end{itemize}

Let $\Pr[\mbox{Type-1}]$ and $\Pr[\mbox{Type-2}]$ denote the probabilities 
with which $\mathcal{A}$  successes as Type-1 forger and Type-2 forger, respectively.
By the proof in Section \ref{prooftheorem:3}, we have $\Pr[\mbox{Type-I}]=\frac{1}{2}\times 
\epsilon_{1}(\frac{p-q}{p}+\frac{1}{p}+\frac{p-1}{p^{3}})$.  Hence, 
$\mathcal{B}$ can break the JoC-$q$-SDH assumption with the advantage 
$\frac{\epsilon_{1}(\ell)}{2}(\frac{p-q}{p}+\frac{1}{p}+\frac{p-1}{p^{3}})$ or 
break the DL assumption with the advantage $\frac{\epsilon_{2}(\ell)}{2}$. 
Therefore, 
$\epsilon(\ell)=max\left\{\frac{\epsilon_{1}(\ell)}{2}(\frac{p-q}{p}+ 
\frac{1}{p}+\frac{p-1}{p^{3}}),\frac{\epsilon_{2}(\ell)}{2}\right\}$.
\end{IEEEproof}

\section{Benchmarking}\label{perf}  
\label{sec:benchmarks}

In this section we evaluate the performance of our scheme. The source code of
the scheme's implementation is available at \cite{githubrepo} and its
performance has been measured on a Dell Inspiron Latitude E5270 laptop with an
Intel Core i7-6600U CPU, 1TB SSD and 16GB of RAM running Fedora 28. %
The implementation makes use of bilinear maps defined over elliptic curves as
well as other cryptographic primitives. For the bilinear maps, we used the JPBC
library\cite{jpbc} wrapper for the C-based implementation of the PBC
libraries~(\cite{lynn:2006}) while bouncycastle\cite{bouncycastle} provides the
other cryptographic primitives required by our scheme.  Note that the implementation by Han \etal\cite{hcsts:asso2018}
 was baed on a Java implementation.

Recall from Section~\ref{preli} that our scheme requires a Type III bilinear
map, $e:\mathbb{G}_1 \times \mathbb{G}_2 \rightarrow \mathbb{G}_{\tau}$. The PBC
library \cite{lynn:2006} provides such an instances in the form of the ``Type
F'' pairing which is based on the  elliptic curve $E:y^{2}=x^{3}+b$. The order
of groups $\mathbb{G}_1$, $\mathbb{G}_2$ and $\mathbb{G}_{\tau}$ is determined by
the group of points on the elliptic curve, $\#E(F_q)=p$.  Note that the Type F
curve is a pairing-friendly Barreto-Naehrig(BN) elliptic
curve~\cite{barreto2005pairing}. In our implementation, we instantiated the Type
F curve using $rBits=256$ and $rBits=638$ where $rBits$ indicates the number of
bits needed to represent the prime $p$.

These bit sizes were chosen to follow the default values specified in the
ECC-DAA standard~\cite{camenischfido} for these curves. Notably, there have 
been recent attacks~\cite{barbulescu2017updating,kim2016extended} against BN 
curves which reduced of the security of an implementation based on the 
$256$-bit curve. However, $638$-bit curve is still considered to be secure.





For the hash functions $H_1:\{0,1\}^{*}\rightarrow\mathbb{Z}_{p}$,
$H_2:\{0,1\}^{*}\rightarrow\mathbb{G}_2$ and {$H_{3}:\{0,1\}^{*}\rightarrow
\{0,1\}^{\ell'} ~ (\ell'<\ell)$ required by our scheme (see Fig~\ref{fig.setup}), we
used $SHA-256$ for both $H_1$ and $H_3$ while for $H_2$ (the random oracle hash
function), we used $SHA-256$ and the ``newRandomElementfromHash()" method in the
JPBC library to construct an element of $\mathbb{G}_2$.


\subsection{Timings} 

Table~\ref{tab:Benchmarks} shows the results of the computational time spent in
those phases of our scheme which required more complex computations
(\ie some form of verification using bilinear maps or generation of zero
knowledge proofs). The timings shown have been calculated as the average over
$50$ iterations.

\inserttabd{tab:Benchmarks}{|l|c|c|c|}
{Benchmark results {(in ms)}} 
{
Protocol phase &Entity& $rBits=256$ &$rBits=638$\\
\hline
\multicolumn{4}{|c|}{Set-up -  Central Authority 
($\mathcal{CA}$)}\\
\hline
initialise the system &CA &$233$ &$971$ \\
\hline
\multicolumn{4}{|c|}{Registration - Issuer ($\mathcal{I}$)}\\
\hline
generate credentials &CA&$3$ &$14$\\
\hline
verify credentials &Issuer&$53$ &$280$ \\
\hline
\multicolumn{4}{|c|}{Registration - User ($\mathcal{U}$)}\\
\hline
generate credentials &CA&$3$ &$14$\\
\hline
verify credentials &User&$52$ &$266$ \\
\hline
\multicolumn{4}{|c|}{Registration - Verifiers ($\mathcal{V}$)}\\
\hline
generate credentials &CA&$6$ &$28$\\
\hline
verify credentials &Verifier&$101$ &$520$ \\
\hline
\multicolumn{4}{|c|}{Registration - Central Verifier ($\mathcal{CV}$)}\\
\hline
generate credentials &CA&$8$ &$41$\\
\hline
verify credentials &Central Verifier&$150$ &$782$ \\
\hline
\multicolumn{4}{|c|}{Ticket Issuing ($4$ services $+$ CV $=5$ tags)}\\
\hline
generate ticket request &User &$44$ &$195$ \\
\hline
generate ticket &Issuer &$291$ &$1432$ \\
\hline
verify ticket &User &$311$ &$1616$ \\
\hline
\multicolumn{4}{|c|}{Ticket Validation - Verifier ($\mathcal{V}$)}\\
\hline
send tag \& proof &User &$4$ &$20$ \\ 
\hline
verify proof \& tag & Verifier &$81$&$421$ \\
\hline
\multicolumn{4}{|c|}{Proxy Verification - Proxy Verifier ($\mathcal{V'}$)}\\
\hline
generate re-key & CA &$5$&$22$ \\
\hline
verify proof \& tag & Proxy Validation &$105$&$545$ \\
\hline
\multicolumn{4}{|c|}{Ticket Trace (5 tags) - Central Verifier 
($\mathcal{CV}$)}\\
\hline
Send ticket \& proof  & User &$4$&$20$ \\
\hline
verify proof \& trace ticket & Central Verifier &$402$&$2083$ \\
\hline
}
The set-up phase is a one off process run by the $\mathcal{CA}$ and only takes 
$233$ms for $rBits=256$ or $971$ms for $rBits=638$.

During the registration phase of the protocol, the generation of credentials by
the $\mathcal{CA}$  for the central verifier, $\mathcal{CV}$, takes the most
computational effort ($8$ms and $42$ms) as this involves the creation of two
credentials. The first one is equivalent to a user credential while the other
one is that of a normal verifier. Similarly, the verification of the
$\mathcal{CV}$'s credentials requires the most computational effort ($150$ms or
$782$ms for $256$ bits and $638$ bits respectively). Note that because of this,
it is unsurprising that the timings for a user and a verifier in this phase add
up to almost the exact number for the $\mathcal{CV}$.%

The ticket issuing phase of our implementation is also reasonably fast when
$rBits=256$. For example, when requesting $4$ services, the whole process takes
$\approx 646$ms, $44$ms to generate the request, $291$ms to produce the ticket
and $311$ms to verify that the ticket is valid. Even when increasing the field
size to $638$ bits, the whole issuing process now takes $\approx 3243$ms of
which $1432$ms is spent on the actual ticket generation by the issuer. Note that
a user can pre-compute her ticket request thus shortening the interaction with
the issuer by $44$ms or $195$ms for $256$-bits and $638$-bits respectively). The
issuer, on the other hand, can also pre-compute some values as part of the
ticket issuing process (\eg $D_V$, $E_V^1$ , $E_V^2$, $K_V$ and parts of $Z_V$,
\cf Fig. \ref{fig.t-i}). This can reduce the ticket issuing phase by another
$193$ms or $965$ms for $256$-bits and $638$-bits respectively.

In the validation phase, verifying an individual tag by a designated verifier
only takes $\approx 85$ms or $\approx 441$ms for $rBits=256$  and $rBits=638$
respectively while acting as a proxy verifier takes slightly longer ($105$ms or
$545$ms) due to the required re-keying of the provided tag. Note, however, that
the generation of the re-key by the $\mathcal{CA}$ is fast ($5$ms or $22$ms).%

Evaluating the performance of a scheme is important to demonstrate its
viability. For example, in the UK, Transport for London(TfL) \cite{Mastercard}
has a requirement for the verification of contactless payment cards used for
travelling to be below 500ms  in order to avoid congestion at ticket barriers.
Given the above performance figures and ignoring any latency introduced by the
communication channel, our ASSO with proxy re-verification scheme is well below
this requirement for $rBits=256$. For $rBits=638$, only the proxy
re-verification is slightly slower ($545$ms) than the required $500$ms. However,
as our implementation has not been optimised for any specific elliptic curve,
additional improvements in speed should be possible.

\section{Conclusions}
\label{sec:conc}

In this paper, a new ASSO with proxy re-verification scheme is proposed which
protects users' privacy and allows a user to authenticate herself to a
designated verifier anonymously. A central authority can authorise new
verifiers to authenticate the user in cases where proxy verification is needed.
The re-key enables the proxy verifier to verify tickets on behalf of the
original ticket verifier on the specified day. However, the proxy verifier
cannot use the re-key to verify tickets with different travel days on behalf of
the original verifier. Furthermore, our scheme is formally treated in terms of
definition, security model and security proof and its performance has been
empirically evaluated.

This work represents one more step in the direction of defining a scheme that
provides strong security and privacy properties in the context of smart
ticketing. We constructed our scheme using the most efficient
pairing (Type-III pairing) available, but the computation cost and communication
cost  may be not suitable for portable devices, e.g. mobile phone, smart card,
tablet, {\em etc.} Further research is needed to optimise the scheme's
construction to minimise the use of pairings in order to potentially improve the
efficiency of the scheme and the associated performance of the implementation in
order to align to the requirements for verification of contactless payment cards
\cite{Mastercard}. An alternative approach to improve
performance is  to construct an ASSO with proxy re-verification scheme which
does not rely on bilinear groups and this is an interesting area of future work.

The rail industry is particularly focused on addressing problems associated with
disruption and the issues surrounding the sharing of passenger details.
Addressing these two concerns is challenging because of the separation of
information held by third party retailers and rail service providers. Retailers
know about passenger travel information but do not necessarily know about events
likely to affect the journey whereas rail service providers know about service
disruptions but do not necessarily have the details to contact the passengers
who might be affected and hence cannot warn them.  Our future research
directions will explore using the techniques presented in this paper to
facilitate privacy-preserving sharing of passenger details between different
parties in the event of disruptions.

\section*{Acknowledgments}
This work has been supported by the EPSRC
Project DICE: ``Data to Improve the Customer Experience'', EP/N028295/1. %
 


\renewcommand{\thesection}{\Alph{section}.\arabic{section}}
\setcounter{section}{0}

\begin{appendices}\label{app}
\section{The Details of Zero-Knowledge Proof}\label{app_zkp}

\subsection{The Detail of $\prod_{U}^{1}$ }\label{d1}

An instantiation of the proof $\prod_{U}^{1}$ is given as follows. 
$\mathcal{U}$ select $y_{1},y_{2},y_{3},x'_{u},e_{u}',y',{y}'_{1},{y}'_{2},k'_{1},k'_{2},\cdots,k'_{|J_{U}|}\stackrel{R}{\leftarrow}\mathbb{Z}_{p}$ and computes $y_{4}=\frac{1}{y_{1}}$,  $y=d_{u}-y_{2}y_{4}$,  $\bar{\sigma}_{U}=\sigma_{U}^{y_{1}}$,  $\tilde{\sigma}_{U}=\bar{\sigma}_{U}^{-e_{u}}A_{U}^{y_{1}}(=\bar{\sigma}_{U}^{x_{a}})$,  $\bar{A}_{U}=A_{U}^{y_{1}}h^{-y_{2}}$,  $\bar{W}_{1}=\bar{\sigma}_{U}^{-e'_{u}}g_{2}^{y'_{1}}$, $\bar{W}_{2}=\bar{A}_{U}^{-y'_{2}}\tilde{g}^{x'_{u}}g_{2}^{y'}$,   $\big(k_{v}=H_{1}(z_{u}||ID_{V}),$  $P_{V}=Y_{U}{Y}_{CV}^{k_{v}}, P'_{v}=\tilde{g}^{x'_{u}}{Y}_{CV}^{k'_{v}}, Q_{V}=\tilde{g}^{k_{v}}, Q'_{V}=\tilde{g}^{k'_{v}}\big)_{ID_{V}\in J_{U}}$. $\mathcal{U}$ computes $c=H_{1}(\bar{\sigma}_{U}||\tilde{\sigma}_{U}||\bar{B}_{U}||\bar{W}_{1}||\bar{W}_{2}||P_{1}||P_{1}'||Q_{1}||Q'_{1}||$ $P_{2}||P_{2}'||Q_{2}||Q'_{2}||\cdots||P_{|J_{U}|}||P_{|J_{U}|}'||$ $Q_{|J_{U}|}||Q'_{|J_{U}|})$,  
$\hat{e}_{u}=e'_{u}-ce_{u}$, $\hat{y}_{1}=y'-cy$, $\hat{y}_{2}=y'_{1}-cy_{2}$, $\hat{y}_{3}=y'_{2}-cy_{4}$,  $\hat{x}_{u}=x'_{u}-cx_{u}$ and $(\hat{k}_{v}=k'_{v}-ck_{v})_{ID_{V}\in J_{U}}$.
 $\mathcal{U}$ sends $(\bar{\sigma}_{U},\tilde{\sigma}_{U},\bar{B}_{U},W_{1},W_{2},(P_{V},P'_{V},Q_{V},Q'_{V})_{ID_{V}\in J_{U}})$ and $(c,\hat{e}_{u},\hat{y}_{1},\hat{y}_{2},\hat{y}_{3},\hat{x}_{u},\hat{k}_{1},\hat{k}_{2},\cdots, \hat{k}_{|J_{U}|})$ to $\mathcal{S}$.
 \medskip
 
After receiving 
$(\bar{\sigma}_{U},\tilde{\sigma}_{U},\bar{A}_{U},\bar{W}_{1},\bar{W}_{2},(P_{V},P'_{V},Q_{V},$ $Q'_{V})_{ID_{V}\in
 J_{U}})$ and 
$(c,\hat{e}_{u},\hat{y}_{1},\hat{y}_{2},\hat{y}_{3},\hat{x}_{u},\hat{k}_{1},\hat{k}_{2},\cdots, \hat{k}_{|J_{U}|})$, $\mathcal{S}$ checks
\begin{equation*}
\begin{split}
 c  \stackrel{?}{=}   & 
H_{1}(\bar{\sigma}_{U}||\tilde{\sigma}_{U}||\bar{A}_{U}||\bar{W}_{1}||\bar{W}_{2}||P_{1}||P_{1}'||Q_{1}||Q'_{1}||P_{2}||P_{2}'||Q_{2}\\
&||Q'_{2}||\cdots||P_{J_{U}}||P_{J_{U}}'||Q_{J_{U}}||Q'_{J_{U}}),
 \end{split}
 \end{equation*}

 \begin{equation*}
   \begin{split}
\bar{W}_{1}\stackrel{?}{=}  
\bar{\sigma}_{U}^{-\hat{e}_{u}}g_{2}^{\hat{v}_{2}}(\frac{\tilde{\sigma}_{U}}{\bar{A}_{U}})^{c},
 ~ \bar{W}_{2}\stackrel{?}{=}  
\bar{A}_{U}^{-\hat{y}_{3}}\tilde{g}^{\hat{x}_{u}}g_{2}^{\hat{y}_{1}}g_{1}^{-c},\\  ~ 
(P'_{V}\stackrel{?}{=}   \tilde{g}^{\hat{x}_{u}}Y_{CV}^{\hat{k}_{v}}P_{V}^{c}, ~
 Q'_{V}\stackrel{?}{=}   \tilde{g}^{\hat{k}_{v}} Q_{V}^{c})_{ID_{V}\in J_{U}}.
 \end{split}
 \end{equation*}

\subsection{The Detail of Zero-Knowledge Proof $\prod_{U}^{2}$ }\label{d2}
An instantiation of the proof $\prod_{U}^{2}$ is as follows. $\mathcal{U}$ selects $x'_{u},k'_{v}\stackrel{R}{\leftarrow}\mathbb{Z}_{p}$, and computes $P_{V}'=\tilde{g}^{x'_{u}}Y_{CV}^{k'_{v}}$, $Q'_{V}=\tilde{g}^{k'_{v}}$, $c_{v}=H_{1}(P_{V}||P'_{V}||Q_{V}||Q'_{V})$, $\hat{x}_{u}=x'_{u}-c_{v}x_{u}$ and $\hat{k}_{v}=k'_{v}-c_{v}k_{v}$. $\mathcal{U}$ sends $(P_{V},P'_{V},Q_{V},Q'_{V})$ and $(c_{v},\hat{x}_{v},\hat{k}_{v})$ to $\mathcal{V}$.
\medskip

After receiving $(P_{V},P'_{V},Q_{V},Q'_{V})$ and $(c_{v},\hat{x}_{v},\hat{k}_{v})$, $\mathcal{V}$ verifiers 
\begin{equation*}
\begin{split}
& c_{v}\stackrel{?}{=}H_{1}(P_{V}||P'_{V}||Q_{V}||Q'_{V}), ~P'_{V}\stackrel{?}{=}\tilde{g}^{\hat{x}_{u}}Y_{CV}^{\hat{k}_{v}}P_{V}^{c_{v}}~\mbox{and}\\
& Q'_{V}\stackrel{?}{=}\tilde{g}^{\hat{k}_{v}}Q_{V}^{c_{v}}.
\end{split}
\end{equation*}

\end{appendices}

\end{document}